\documentclass[12pt]{article}
\usepackage{titlesec}
\usepackage{lipsum}
\usepackage{lineno}
\usepackage{changepage}
\usepackage{graphicx}
\usepackage{amsmath}
\usepackage{amssymb}
\usepackage{caption}
\usepackage{setspace}
\usepackage{color}
\usepackage{comment}
\usepackage{authblk}
\usepackage{bm}
\usepackage{textcomp}
\usepackage{url}

\usepackage[numbers,sort&compress]{natbib}

\usepackage{newtxtext,newtxmath}

\usepackage[
  left=3cm,    
  right=2.6cm,   
  top=2.1cm,     
  bottom=2.1cm,  
]{geometry}

\definecolor{qired}{rgb}{0.6, 0, 0}

\definecolor{tealblue}{rgb}{0.0, 0.5, 0.5}

\titleformat{\section}
{\Large\bfseries}{\thesection}{1em}{}

\DeclareCaptionFormat{boldnumberfirstsentence}{
  \textbf{#1#2}#3\par
}

\DeclareCaptionLabelSeparator{colonspace}{: }
\newcommand*\patchAmsMathEnvironmentForLineno[1]{%
  \expandafter\let\csname old#1\expandafter\endcsname\csname #1\endcsname
  \expandafter\let\csname oldend#1\expandafter\endcsname\csname end#1\endcsname
  \renewenvironment{#1}%
     {\linenomath\csname old#1\endcsname}%
     {\csname oldend#1\endcsname\endlinenomath}}%
\newcommand*\patchBothAmsMathEnvironmentsForLineno[1]{%
  \patchAmsMathEnvironmentForLineno{#1}%
  \patchAmsMathEnvironmentForLineno{#1*}}%
\AtBeginDocument{%
\patchBothAmsMathEnvironmentsForLineno{equation}%
\patchBothAmsMathEnvironmentsForLineno{align}%
\patchBothAmsMathEnvironmentsForLineno{flalign}%
\patchBothAmsMathEnvironmentsForLineno{alignat}%
\patchBothAmsMathEnvironmentsForLineno{gather}%
\patchBothAmsMathEnvironmentsForLineno{multline}%
}

\title{\begin{center}  \bfseries \singlespacing
Fostering Sustainable Cooperation through Strategic Resource Allocation and Utilization on Social Networks
\end{center}
\date{}}
\author{\parbox[c]{16cm}{\centering 
Juyi Li$^{1}$ \quad 
Xiaoqun Wu$^{2}\footnote{Correspondence: 
xqwu@whu.edu.cn}$ \quad 
Qi Su$^{1}\footnote{Correspondence: qisu@sjtu.edu.cn}$ \\ 
		\footnotesize
  $^{1}$School of Automation and Intelligent Sensing, Shanghai Jiao Tong University, Shanghai, 200240, China \\
  $^{2}$School of Computer Science and Software Engineering, Shenzhen University, Guangdong, 518060, China \\
  }
}

\begin{document}
\maketitle
\vspace{-3em}

\begingroup
\titleformat*{\section}{\normalfont\large\bfseries\centering}
\section*{Abstract}
\begin{adjustwidth}{1cm}{1cm} 
\small 

Efficient allocation and use of limited resources are fundamental to advancing collective welfare and achieving long-term societal sustainability. This challenge involves not only how policymakers distribute scarce resources among individuals, but also how individuals strategically utilize them. The complexity deepens when individuals are embedded in networks of social interactions, where outcomes are interdependent and future decisions are shaped by a dynamic tension between cooperation driven by collective long-term benefit and self-interest motivated by short-term personal gain. Here, we introduce a novel framework of generalized public goods games on hypergraphs to capture the multifaceted nature of real-world social interactions. Using Nash equilibrium analysis, we reveal how full cooperation (all individuals contribute all their resources to maximize collective benefit) emerges from the interplay between resource allocation strategies, individual usage behaviors, and the structure of interactions. We find that equal resource distribution enhances cooperation in homogeneous networks but may suppress it in heterogeneous ones, indicating that equity in allocation does not universally lead to optimal collective outcomes. To address this, we propose two complementary optimization strategies: one to guide policymakers in designing effective resource allocation schemes, and the other to support individuals in making sustainable use decisions. We validate the effectiveness of both approaches across a range of synthetic and empirical cases. {Moreover, using real-world data from three sustainability domains (climate commons, water sharing, and renewable resource management), we find that both approaches are broadly applicable for promoting full cooperation.} Our findings provide actionable insights for designing governance frameworks and resource management policies that promote sustainable cooperation in complex socio-environmental systems.

\end{adjustwidth}
\endgroup

\section*{Introduction}

Cooperation is essential for tackling some of the most pressing sustainability challenges of our time, from mitigating climate change and managing shared natural resources to coordinating global responses to public health crises. Achieving such cooperation often requires individuals, organizations, and nations to make short-term sacrifices for long-term collective benefits. Yet these efforts are vulnerable to self-interested behaviors that prioritize immediate gains over shared prosperity, creating social dilemmas that undermine collective action \cite{ref1}. While evolutionary game theory has offered foundational insights into these dynamics, much of the traditional focus has been on pairwise interactions, exemplified by models such as the prisoner’s dilemma \cite{rapoport1965}, hawk-dove game \cite{maynard-smith1973}, snowdrift game \cite{doebeli2005}, and stag hunt game \cite{rousseau1755}. However, real-world cooperation often arises in more complex, multi-player settings that cannot be reduced to a sum of pairwise interactions. For instance, the Paris Climate Agreement \cite{Paris} exemplifies a global coordination effort in which multiple actors must weigh short-term economic sacrifices against long-term environmental gains. Similarly, decision-making in the UN Security Council \cite{Voeten_2001} involves strategic negotiation shaped by institutional rules and conflicting national interests, while fiscal policy deliberations within the European Union \cite{European}  illustrate how consensus emerges through coalition-building and shifting alignments among states. These examples highlight the significance of higher-order interactions in shaping cooperative outcomes. Unlike dyadic games, multi-player frameworks reveal a broader and more nuanced strategy landscape, offering deeper insights into the dynamics of collaboration, conflict, and coordination in complex social systems. Crucially, the structural complexity of multi-player interactions cannot be fully captured by simple linear combinations of pairwise games.

{The public goods game is a foundational framework for studying multi-player interactions and collective behavior \cite{ref37,ref40,ref42,ref43}, particularly in understanding mechanisms of resource allocation and incentive design \cite{chen2014optimal,perc2017statistical}.} It mirrors many real-world challenges, such as community resource management, infrastructure development, and climate mitigation efforts, where individuals must decide how much of their resources to contribute toward shared benefits. In its classical formulation, individuals are endowed with equal initial resources and must decide whether to contribute to a shared pool. Contributions are then multiplied by a productivity factor and evenly redistributed among all participants, regardless of individual input \cite{isaac1988group,su2019spatial}.
This simple yet powerful setup captures the tension between personal incentives and group welfare that lies at the heart of many sustainability dilemmas.
{Despite its widespread use, much of the existing literature assumes a homogeneous setting \cite{ref12,ref13,ref14,ref15}, thereby neglecting the diverse forms of heterogeneity commonly observed in real-world systems \cite{su2018understanding,perc2011success,perc2011does}.
These include: (1) endowment inequality, where economic disparities result in players receiving unequal initial resources \cite{ref16}; (2) productivity inequality, often arising from differential access to education or technology, leading to variation in individual productivity levels \cite{ref17}; (3) contribution inequality, where individuals differ in their willingness and capacity to contribute to public goods, as seen in contexts like infrastructure development \cite{janssen2011coordination} or charitable activities \cite{auerbach2013handbook}.}
In such settings, strategic choices are no longer confined to binary decisions but span a continuous and context-dependent spectrum. Understanding how these layered heterogeneities shape cooperative dynamics remains a pressing challenge in the study of collective behavior.

Repeated interactions are a defining feature of real-world social systems and play a critical role in sustaining cooperation over time. For example, individuals often engage in long-term community environmental initiatives \cite{anderson2024cooperation}, where ongoing collaboration is essential to preserving shared resources, or participate in enduring business partnerships \cite{frey2019long}, where mutual trust is built through repeated cooperation.
Moreover, individuals are rarely limited to a single interaction; instead, they typically take part in multiple concurrent group activities. Such patterns are evident across diverse contexts, such as researchers collaborating with multiple teams \cite{pan2012world}, residents contributing to several community projects \cite{centola2007complex}, and users participating in various online platforms \cite{jaidka2022cross}.
Studying systems where individuals engage in repeated interactions across multiple groups not only better captures the complexity of real-world social structures but also introduces significant analytical challenges. These features make such systems both highly realistic and theoretically rich, underscoring the need for deeper investigation.

Accurately capturing overlapping and repeated group participation is essential for a deeper understanding of collective behavior. In this regard, networked game theory has emerged as a powerful framework. Most existing studies have centered on pairwise interactions, where individuals engage with their immediate neighbors. These models have been successfully extended to encompass complex network structures \cite{ref3,mcavoy2021fixation,mcavoy2020social,ref2}, temporal dynamics \cite{ref64,li2020evolution}, and multilayer configurations \cite{ref45}, yielding valuable insights into the evolution of cooperation.
However, the inherently dyadic structure of such models presents a fundamental limitation: they are poorly equipped to represent the group-level mechanisms that underpin many real-world collective behaviors. {To overcome this, researchers have increasingly adopted higher-order interaction models \cite{ref22,ref23,ref24,ref25,ref26,ma2024social}, using advanced mathematical tools such as hypergraphs \cite{ref8,ref30,ref31} and simplicial complexes \cite{ref7,ref32,ref33,ref34,xu2024reinforcement}, which enable explicit modeling of multi-player relationships and overlapping group memberships.
Among these, hypergraphs are particularly well-suited for capturing the structure of repeated public goods games, especially when individuals belong to multiple potentially asymmetric groups \cite{ref9,xu2024reinforcement}.} Despite their promise, current studies on hypergraph-based games often rest on assumptions of homogeneity or employ mean-field approximations \cite{ref52,ref8}, limiting their relevance and applicability in heterogeneous, real-world contexts.

In this work, we propose a general framework of iterated multi-player public goods games on hypergraphs to study repeated multi-player interactions in structured populations.
In our model, each individual may belong to multiple hyperedges, with individuals within each hyperedge engaging in an iterated public goods game.
{We introduce a measure of cooperation termed ``full cooperation,'' defined as a state where all players contribute their entire endowments to maximize collective benefits; for instance, in transboundary water governance, full cooperation corresponds to each country fully contributing its agreed GDP-based share to joint river management, without withholding or under-providing relative to its commitment. We derive necessary and sufficient conditions for achieving full cooperation in structured populations, applicable to arbitrary hypergraph structures. Through Nash equilibrium analysis, we uncover the complex relationships linking full cooperation to key determinants, including initial endowments, productivity levels, individual contribution strategies, and interaction structure. Remarkably, equal endowments promote full cooperation most effectively in homogeneous hypergraphs, yet may hinder cooperation in heterogeneous hypergraphs, highlighting that equal endowments are not always the optimal solution. To address this challenge, we propose a dual-intervention mechanism involving differentiated allocation policies implemented by regulatory authorities and individual optimization of contribution strategies by participants. Experimental validation confirms that both mechanisms effectively foster full cooperation in large-scale hypergraphs and empirical hypergraphs. In addition, tests across three sustainability domains show that both mechanisms are broadly present and promote full cooperation.} These findings offer actionable insights for governance frameworks and resource management policies seeking to enhance cooperation in complex, interconnected societies.

\begin{figure}
    \centering
    \includegraphics[width=\linewidth]{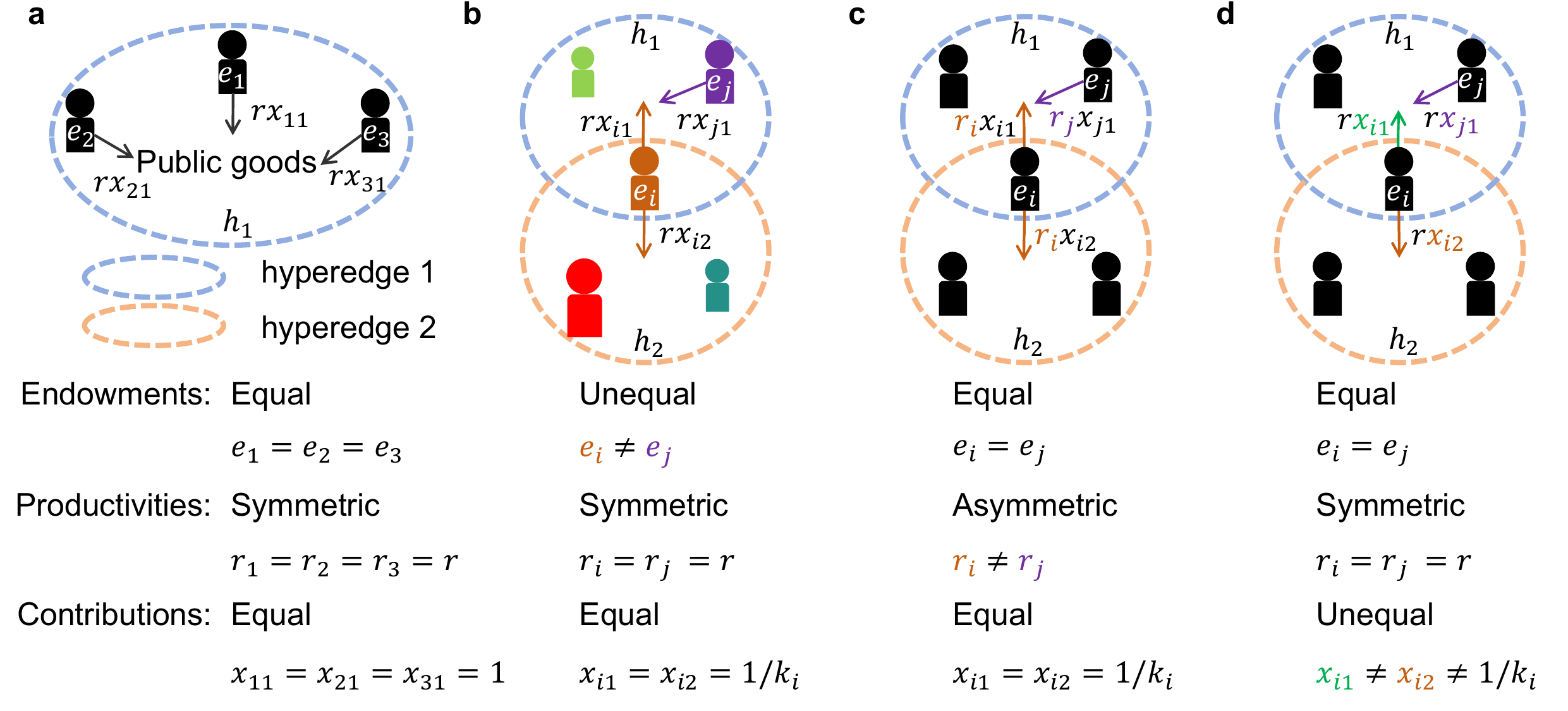}
  \caption{\textbf{Public goods games on hypergraphs.} 
    We investigate repeated public goods games in two settings: a single-group scenario (\textbf{a}) and a hypergraph framework with multiple groups (e.g., a two-group hypergraph in \textbf{bcd}). 
    In each group, players engage in a repeated public goods game. Some players may participate in multiple games if they belong to multiple groups.
    In each round, player $i$ receives an initial endowment $e_i$ and decides what fraction of their endowment to contribute to each group they are part of. For example, player $i$ may allocate a fraction $x_{i1}$ to group 1.
    The contributed amount is then multiplied by the player's productivity $r_i$, and the total is evenly distributed among all participants in that group.
    The outcome of the public goods game is determined by three key factors: each player's initial endowment $e_i$, productivity $r_i$, and their contribution $x_{ih}$ to each group. 
    \textbf{a}, The baseline model of a public goods game in a single-group scenario, where all players have identical endowments and productivity, contributing the maximum amount to the single group.
    \textbf{b}-\textbf{d} illustrate public goods games on hypergraphs with multiple groups, where five players participate in two distinct games, labeled $h_1$ and $h_2$.
    Player $i$ participates in all two games.
    \textbf{b}, Public goods games on hypergraphs with symmetric productivity and equal contributions, but unequal endowments $e_i\neq e_j $. 
    \textbf{c}, Public goods games on hypergraphs with equal endowments and contributions, but asymmetric productivity $r_i\neq r_j$.
    \textbf{d}, Public goods games on hypergraphs with equal endowments and symmetric productivity, but unequal contributions $x_{i1}\neq x_{i2}$.}
    \label{fig1}
\end{figure}

\section*{Model}

We consider a population of \( N \) players, denoted by \( \mathcal{N} = \{1, 2, \dots, N\} \), engaged in \( M \) iterated games, indexed by \( \mathcal{M} = \{1, 2, \dots, M\} \). Each player may participate in multiple games, and interactions are repeated over time with a continuation probability \( \delta \), representing the likelihood of proceeding to the next round after one iteration concludes. Our focus is on public goods games, which are repeatedly played within different groups of players.
To model these structured group interactions, we adopt a hypergraph representation~\cite{ref26,ref31}, where nodes correspond to players and hyperedges represent public goods games. The participation structure is encoded by an incidence matrix \( A = \{a_{ij}\} \in \mathbb{R}^{N \times M} \), where \( a_{ij} = 1 \) if player \( i \) participates in game \( j \), and \( a_{ij} = 0 \) otherwise. We assume that each hyperedge (i.e., each public goods game) has a fixed size \( \sigma \), ensuring that all games involve the same number of participants.

In each round of the game, player \( i \) receives a fixed endowment denoted by \( e_i \), which can be interpreted as a regular income. The full distribution of endowments across the population is represented by the vector \( \boldsymbol{e} = \{e_1, e_2, \dots, e_N\} \). Without loss of generality, we assume that this vector is normalized such that \( \sum_{i=1}^{N} e_i = 1 \), ensuring a fixed total resource pool.
A special case is the equal endowment scenario, given by \( \boldsymbol{e} = \left\{\frac{1}{N}, \frac{1}{N}, \dots, \frac{1}{N} \right\} \), where each player receives an identical share of the total resources. By contrast, the vector \( \boldsymbol{e} = \{1, 0, \dots, 0\} \) represents an extreme form of inequality, where the first player receives all available endowments and the remaining players receive none.

Upon receiving their endowment \( e_i \), each player \( i \) decides how to allocate it across the public goods games in which they participate. This decision is represented by the contribution matrix \( X = \{x_{ij}\} \in \mathbb{R}^{N \times M} \), where \( 0 \leq x_{ij} \leq 1 \) denotes the fraction of player \( i \)'s endowment allocated to game \( j \).
Two key constraints govern this allocation process. First, contributions are only defined for games in which player \( i \) participates, i.e., \( x_{ij} > 0 \) only if \( a_{ij} = 1 \). Second, the total fraction of the endowment distributed by each player across all games cannot exceed 1. Formally, this implies \( \sum_{j=1}^{M} x_{ij} \leq 1 \) for all \( i \in \mathcal{N} \). A row sum of 1 indicates that the player fully allocates their endowment, while a sum less than 1 implies that a portion is retained. The condition \( \sum_{j=1}^{M} x_{ij} = 1 \) for all \( i \) corresponds to a state of full cooperation within the population, where every player contributes their entire endowment to the public goods games.

Given an endowment vector \( \boldsymbol{e} \) and a contribution matrix \( X \), each player allocates their endowment across the public goods games in which they participate. In each game, every player's contribution is multiplied by their own productivity coefficient and then equally redistributed among all participants.
Let \( r_j \) denote the productivity coefficient associated with player \( j \), which reflects their effectiveness in generating returns from contributions. The resulting payoff for player \( i \) is given by:
\begin{equation}
u_i(\boldsymbol{e}, X) = \sum_{k=1}^{M} \frac{a_{ik}}{\sigma} \sum_{j=1}^{N} r_j x_{jk} e_j + \left(1 - \sum_{k=1}^{M} x_{ik} \right) e_i,
\label{linear_asymmetric_structure}
\end{equation}
where the first term represents the returns player \( i \) receives from participating in public goods games, and the second term accounts for the portion of their endowment retained for private use.
We classify such payoff functions into two categories: linear symmetric productivity and linear asymmetric productivity. The term ``linear'' refers to the linear dependence of payoffs on both contributions and productivity coefficients. In the symmetric case, all players share a common productivity coefficient \( r_j = r \), whereas in the asymmetric case, each player has an individual, fixed productivity \( r_j \), which may vary across the population.
To ensure the presence of a social dilemma, we assume the productivity coefficients satisfy \( 1 < r_j < \sigma \) for all \( j \in \mathcal{N} \). This range guarantees that while cooperation can benefit the group, individuals are still incentivized to defect in one-shot interactions, corresponding to a social dilemma (see SI Appendix, Section 2.3). If \( r_j \) exceeds \( \sigma \), cooperation becomes inherently advantageous, thus eliminating the tension between individual and collective interests~\cite{ref71}.

In the special case where the system consists of a single hyperedge, productivity is symmetric (\( r_j = r \)), endowments are equally distributed, and players adopt binary contribution strategies (\( x_{ik} \in \{0, 1\} \), representing defection or cooperation), the model reduces to the classical public goods game. In this setting, full cooperation corresponds to the scenario in which all players choose to cooperate (Fig.~\ref{fig1}\textbf{a}).
While the model simplifies under these homogeneous conditions, real-world social dilemmas are often characterized by heterogeneity among individuals, particularly in endowment levels and productivity (Fig.~\ref{fig1}\textbf{b}, \textbf{c}). Such heterogeneity introduces strategic complexities absent from the classical framework. 
Furthermore, when individuals participate in multiple public goods games, their strategic space expands beyond simple binary choices. Unlike the single-game case where full cooperation entails contributing one's entire endowment to a single pool, multi-game participation allows for flexible contribution allocations. For example, a player \( i \) involved in two games may opt for an even split of contributions (\( x_{i1} = 0.5 \), \( x_{i2} = 0.5 \)) (Fig.~\ref{fig1}\textbf{b}, \textbf{c}), or concentrate their contributions entirely in one game (\( x_{i1} = 1 \), \( x_{i2} = 0 \)) (Fig.~\ref{fig1}\textbf{d}). These diverse strategies can substantially influence the emergence and stability of cooperation, highlighting the need for systematic investigation of cooperative dynamics in heterogeneous, multi-game environments.

\section*{Results}

In the following, we first derive the general necessary and sufficient conditions for achieving full cooperation across arbitrary hypergraph structures. We then explore how different endowment distributions influence full cooperation. Next, we investigate the role of contribution strategies in shaping cooperative outcomes. To effectively foster full cooperation, we propose two intervention mechanisms: one involving endowment redistribution policies implemented by policymakers, and the other based on individual optimization of contribution strategies. Finally, we apply and validate these two mechanisms within large-scale hypergraphs and empirical hypergraphs, demonstrating their effectiveness in promoting full cooperation.

\subsection*{General condition for full cooperation}

First, we show that in linear public goods games with a given payoff function and fixed endowment allocation, full cooperation is achievable if there exists a subgame perfect equilibrium in which all players contribute their entire endowments. In such an equilibrium, no player has an incentive to deviate from fully cooperative behavior \cite{ref72}. Moreover, we establish that full cooperation is possible if and only if the Grim strategy constitutes an equilibrium (see SI Appendix, Section 2.1). Under the Grim strategy, player $i$ initially fully cooperates, but if any member of their associated hyperedge defects, player $i$ responds by defecting in all subsequent rounds \cite{ref73}.

For a given initial endowment vector \(\boldsymbol{e} = \{e_1, e_2, \dots, e_N\}\) and a deterministic contribution matrix \( X \), full cooperation is feasible under the Grim equilibrium condition if and only if (SI Appendix, section 2.2):
\begin{equation}
    \begin{split}
        \delta &\geq \frac{(\sigma-r_i)e_i}{\sum_{k=1}^{M}a_{ik}\sum_{j\neq i}^{N}r_jx_{jk}e_j}
    \end{split}
    \label{linear asymmetric sufficient}
\end{equation}
for every player $i$.
If inequality (\ref{linear asymmetric sufficient}) is satisfied, it indicates that the immediate gains from defection for player \( i \) are outweighed by the long-term benefits of maintaining full cooperation. Thus, achieving and sustaining full cooperation within a hypergraph requires that the discounted future benefits for all players exceed any short-term incentives to defect.
A crucial point is that full cooperation demands participation from every individual. Therefore, inequality (\ref{linear asymmetric sufficient}) must be satisfied for each player. If even a single player fails to meet this condition, they would have an incentive to defect, ultimately preventing the group from attaining full cooperation.

Each node $i$ is associated with a unique threshold for the continuation probability of full cooperation, denoted as $\delta_{i}^{*}=\frac{(\sigma-r_i)e_i}{\sum_{k=1}^{M}a_{ik}\sum_{j\neq i}^{N}r_jx_{jk}e_j}$. The feasibility of full cooperation within the hypergraph is contingent upon whether the specified $\delta$ values equal or surpass the maximum threshold requirements for all nodes. Hence, our primary goal is to identify the upper bound $\max\limits_i{\delta_i^{*}}$, as the player corresponding to this bound is likely to be the first to abandon full cooperation. The threshold $\delta_{i}^{*}$ for the node that first defects from full cooperation is characterized by a relatively large numerator and a small denominator. For the given hyperedge dimension $\sigma$, both productivity $r_i$ and individual endowment $e_i$ can influence the numerator. Specifically, a larger numerator occurs when $e_i$ is greater and $r_i$ is smaller. Thus, a player with a higher initial endowment and lower productivity contributes to a larger numerator. In contrast, the denominator depends on the number of hyperedges associated with player $i$, captured by  $a_{ik}$, and the contributions from other players $j$ within each hyperedge $k$ that player $i$ participates in, captured by $r_jx_{jk}e_j$. A smaller denominator indicates that player $i$ is involved in fewer hyperedges or participates in hyperedges where other players contribute relatively less of their endowments.

Building on the above discussion, we infer that individuals most likely to deviate from full cooperation are those who initially receive higher endowments, have lower productivity coefficients, and participate in fewer group interactions. Their relatively high $\delta_i^{*}$ values imply that as the continuation probability $\delta$ gradually decreases from 1, it will first fall below these thresholds. Once this occurs, the condition specified by inequality (\ref{linear asymmetric sufficient}) is no longer satisfied for these individuals, ultimately undermining the system’s ability to sustain full cooperation.

\subsection*{Effects of endowment distribution on full cooperation}

\begin{figure}
    \centering
    \includegraphics[width=\linewidth]{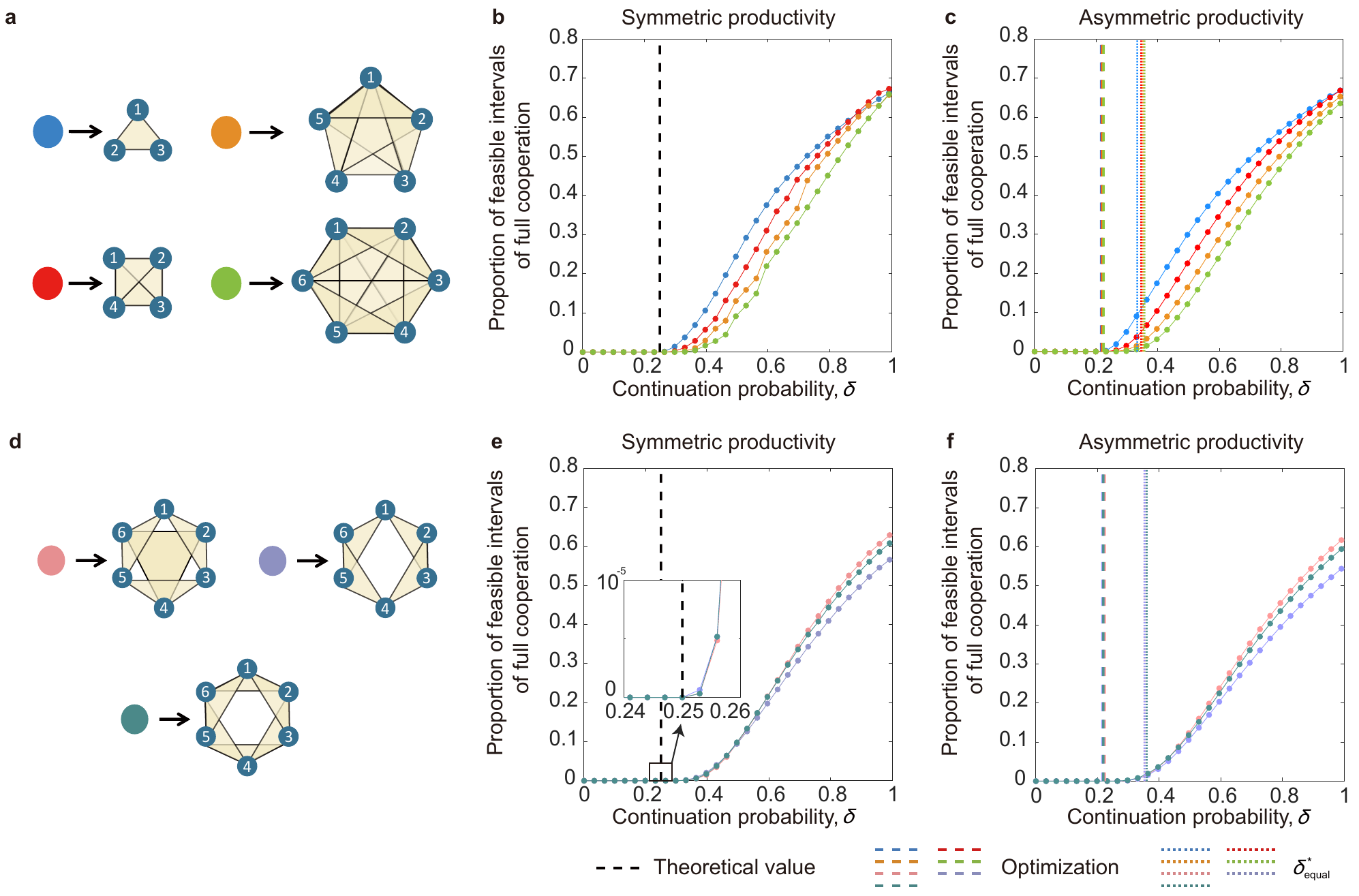}
    \caption{\textbf{Equal endowments foster cooperation under symmetric productivity, while unequal endowments are more effective when productivity is asymmetric.} We consider both fully connected homogeneous hypergraphs (\textbf{a}) and partially connected homogeneous hypergraphs (\textbf{d}). Panels \textbf{b}, \textbf{c}, \textbf{e}, and \textbf{f} display the proportion of endowment combinations leading to the full cooperation, within the entire endowment space \( E \) and for different continuation probability \(\delta\). 
    \textbf{a}, Fully connected hypergraphs with 3 to 6 nodes are color-coded blue, red, yellow, and green, respectively, with three players per game and total game counts of 1, 4, 10, and 20.  \textbf{d}, Partially connected homogeneous hypergraphs with six nodes, pink and purple hyperedges have an average hyperdegree of 2, while dark green hyperedges have an average hyperdegree of 3.  \textbf{b} and \textbf{e} show the proportion of endowment combinations leading to full cooperation within the endowment space \( E \), with symmetric productivity (\( r_i = 2 \)). \textbf{c} and \textbf{f} illustrate the proportion of endowment combinations leading to the full cooperation, with asymmetric productivity (\( r_1 = 1.5 \), \( r_2 = 2 \), \( r_3 = 2.5 \), and \( r_4 = r_5 = r_6 = 2 \)). This setup ensures a consistent average productivity across hypergraphs with varying node counts. Black dashed lines indicate the theoretical optimum \( e_i = \frac{1}{N} \), colored dashed lines represent optimized endowment allocations (see the Materials and Methods section), and dotted lines denote the continuation probability \(\delta_{\text{equal}}^{*}\) required for equal endowments under asymmetric productivity. Overall, \textbf{b} and \textbf{e} confirm that equal endowments best promote cooperation under symmetric productivity, while \textbf{c} and \textbf{f} show that unequal endowments better promote cooperation in the presence of asymmetric productivity.}
    
    \label{fig2}
\end{figure}

We examine how the distribution of endowments influences the emergence of full cooperation, with particular attention to whether equal endowments across individuals are optimal for fostering full cooperation under both symmetric and asymmetric productivity scenarios.
In this section, we focus on an equal-contribution setting, where each player allocates their contributions uniformly across all hyperedges they participate in (unequal contribution strategies are examined in the next section). The endowment space is defined as
\[
E = \left\{ \boldsymbol{e} = (e_1, e_2, \dots, e_N) \;\middle|\; \sum_{i=1}^N e_i = 1,\; e_i \in [0,1] \text{ for all } i \right\}.
\]
Given a continuation probability \(\delta \leq 1\), full cooperation is feasible if there exists at least one endowment vector \((e_1, e_2, \dots, e_N)\) that satisfies inequality~(\ref{linear asymmetric sufficient}). Furthermore, for a fixed \(\delta\), a setup is considered more favorable to full cooperation if it admits a larger number of such feasible endowment combinations.

We consider a fully connected homogeneous hypergraph, in which all possible hyperedges are present. In a hypergraph with \(N\) nodes and hyperedges connecting \(\sigma\) individuals, the total number of hyperedges is \(\binom{N}{\sigma}\), and each node's hyperdegree (i.e., the number of hyperedges a node participates in) is \(\binom{N-1}{\sigma-1}\). For instance, when \(\sigma = 3\), the number of hyperedges is 1, 4, 10, or 20 for \(N = 3, 4, 5,\) or 6, respectively, and the corresponding hyperdegrees are 1, 3, 6, and 10 (Fig.~\ref{fig2}\textbf{a}).
Figure~\ref{fig2}\textbf{b} illustrates that under symmetric productivity—where all individuals have identical productivity rates—increasing the continuation probability \(\delta\) consistently expands the set of endowment combinations that support full cooperation (see the formal proof in SI Appendix, Lemma 2.2).
We are particularly interested in the smallest continuation probability that enables full cooperation for at least one endowment configuration, denoted by \(\delta^\ast\), along with the corresponding endowment vectors. We refer to these vectors as ``optimal endowment combinations'' since they enable full cooperation at the lowest possible continuation probability.
Figure~\ref{fig2}\textbf{b} further reveals that under symmetric productivity, the smallest continuation probability that permits full cooperation with equal endowments, denoted by \(\delta^\ast_{\text{equal}}\), coincides with \(\delta^\ast\). Moreover, we show that the optimal endowment combination in this case is the equal endowment distribution, i.e., \(e_i = 1/N\) for all \(i\).
Thus, we conclude that under symmetric productivity, equal endowments are optimal for fostering cooperation, as they ensure full cooperation over the broadest possible range of continuation probabilities.

We also calculate the continuation probability threshold for equal endowments under the asymmetric productivity payoff function (Fig.~\ref{fig2}\textbf{c}). Our results show that equal endowments are not optimal, i.e., \(\delta_\text{equal}^{*} > \delta^{*}\). That is, for \(\delta \in (\delta^{*}, \delta_\text{equal}^{*})\), equal endowments fails to induce full cooperation but some unequal endowments can make it.

Beyond fully connected hypergraphs, we also investigate partially connected homogeneous hypergraphs. We prove that under a symmetric productivity payoff function, for a given productivity \( r \) and hyperedge dimension \( \sigma \), all homogeneous hypergraph structures share a common smallest continuation probability (SI Appendix, Section 2.6), given by
$\delta^{*} = \frac{\sigma - r}{r(\sigma - 1)}.$
This threshold is independent of the hyperdegree, implying that the smallest continuation probability required for full cooperation remains constant across different network densities. Moreover, we find that \(\delta^{*} = \delta_{\text{equal}}^{*}\), indicating that equal endowments continue to be optimal even in partially connected homogeneous hypergraphs (Fig.~\ref{fig2}\textbf{e}). 
Under asymmetric productivity, however, unequal endowments can further facilitate cooperation (Fig.~\ref{fig2}\textbf{f}). Notably, the expression \(\delta^{*} = \frac{\sigma - r}{r(\sigma - 1)}\) holds not only for small networks but also for large-scale homogeneous hypergraphs. To validate this theoretical result, we conducted extensive simulations on large networks, confirming its accuracy (Extended Data~\ref{supplementary_figure6}).

\subsection*{Effects of contribution adjustments on full cooperation}

\begin{figure}
    \centering
    \includegraphics[width=\linewidth]{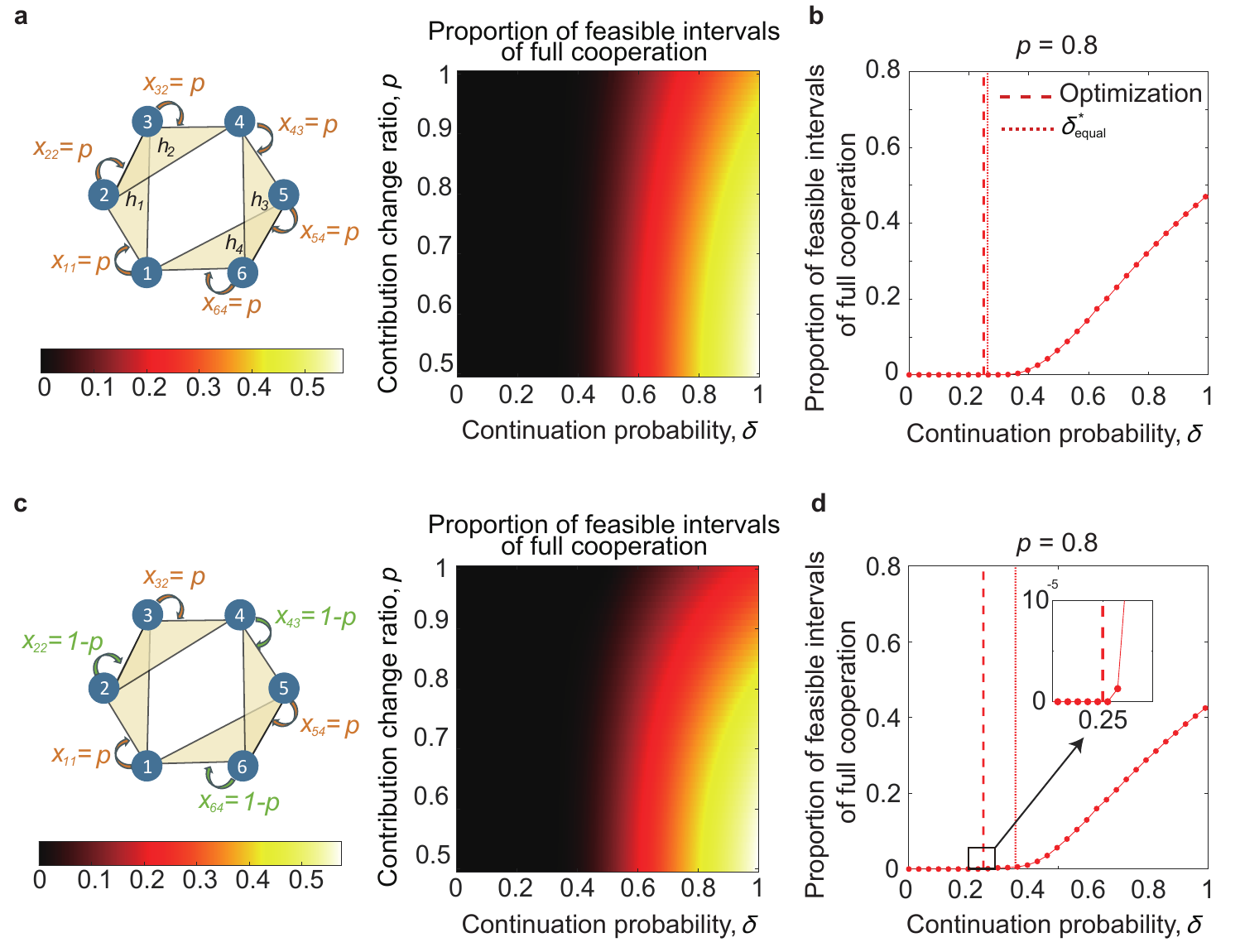}
    \caption{\textbf{Equal contribuions most effectively promote full coopration.}
    We consider a partially connected homogeneous hypergraph with symmetric productivity ($r_i=2$), where players 1, 2, and 3 participate in game \( h_1 \); players 2, 3, and 4 in \( h_2 \); players 4, 5, and 6 in \( h_3 \); and players 5, 6, and 1 in \( h_4 \). 
    \textbf{a}, The impact of contribution proportion \( p \) on the proportion of endowment combinations leading to full cooperation. 
    Each player engages in two games, allocating a fraction of \( p \) of their endowment to one game and \( 1-p \) to the other (e.g., player 1 contributes \( x_{11} = p \) to \( h_1 \) and \( x_{14} = 1-p \) to \( h_4 \)). As \( p \) increases from 0.5 to 1, the proportion of fully cooperative endowment combinations decreases.
    \textbf{b}, The proportion of fully cooperative endowment combinations is shown across a broad range of continuation probabilities \(\delta\), with \( p = 0.8 \) from \textbf{a}. 
    The red dashed line represents the optimized endowments for each \(\delta\), while the dotted line indicates the \(\delta_{\text{equal}}^{*}\) for equal endowments, demonstrating that unequal endowments better facilitate cooperation when players distribute their contributions unevenly across different games.
    \textbf{c}, Contribution deviations are further examined by shifting the contributions of players 1, 3, and 5 to the right and those of players 2, 4, and 6 to the left, revealing that increased deviations further reduce the proportion of cooperative outcomes.
    \textbf{d}, when \( p = 0.8 \),  unequal endowments more effectively promote cooperation. Additionally, inset graphs near the optimized \(\delta\) values validate the accuracy of the optimization algorithm.}   
    \label{fig3}
\end{figure}

So far, we have clarified how the allocation of the initial endowment influences full cooperation when individuals distribute their contributions evenly across all interactions. In this section, we examine how, given a fixed initial endowment, individuals’ contribution strategies across interactions can most effectively foster full cooperation.

We investigate a partially connected homogeneous hypergraph with 6 nodes and a node hyperdegree of 2 (Fig.~\ref{fig3}\textbf{a}), where each individual participates in two three-player repeated public goods games. In this setup, each player contributes a proportion \( p \) to one interaction and \( 1 - p \) to the other. For example, player 1 contributes \( p \) to hyperedge \( h_1 \), player 2 contributes \( p \) to \( h_2 \), player 3 to \( h_2 \), player 4 to \( h_3 \), and players 5 and 6 contribute \( p \) to \( h_4 \). When \( p = 0.5 \), every individual contributes equally to all interactions. Deviations from \( p = 0.5 \) indicate unequal contributions across the games.
Figure~\ref{fig3}\textbf{a} shows that increasing \( p \) from 0.5 to 1 results in a reduction in the proportion of feasible intervals for full cooperation. Further experiments, where contributions are adjusted for players 1, 3, and 5 on one side, and players 2, 4, and 6 on the opposite side, reveal a similar decline in the proportion of feasible intervals for full cooperation (Fig.~\ref{fig3}\textbf{c}). We also explore additional scenarios, including changes in the contributions of one player, two players, and arbitrary contributions across all players (Extended Data \ref{supplementary_figure1}). Regardless of how the contribution direction is modified, equal contributions are consistently found to be the most effective in promoting cooperation in homogeneous hypergraphs.
Moreover, when \( p = 0.8 \) (Fig.~\ref{fig3}\textbf{b}, \textbf{d}), 
We observe that the smallest continuation probability $\delta^{*}$ corresponding to the optimal endowment is lower than $\delta_{\textbf{equal}}^{*}$.
This suggests that when contributions are unequal, unequal endowments are more effective in promoting cooperation.

\begin{figure}
    \centering
    \includegraphics[width=0.85\linewidth]{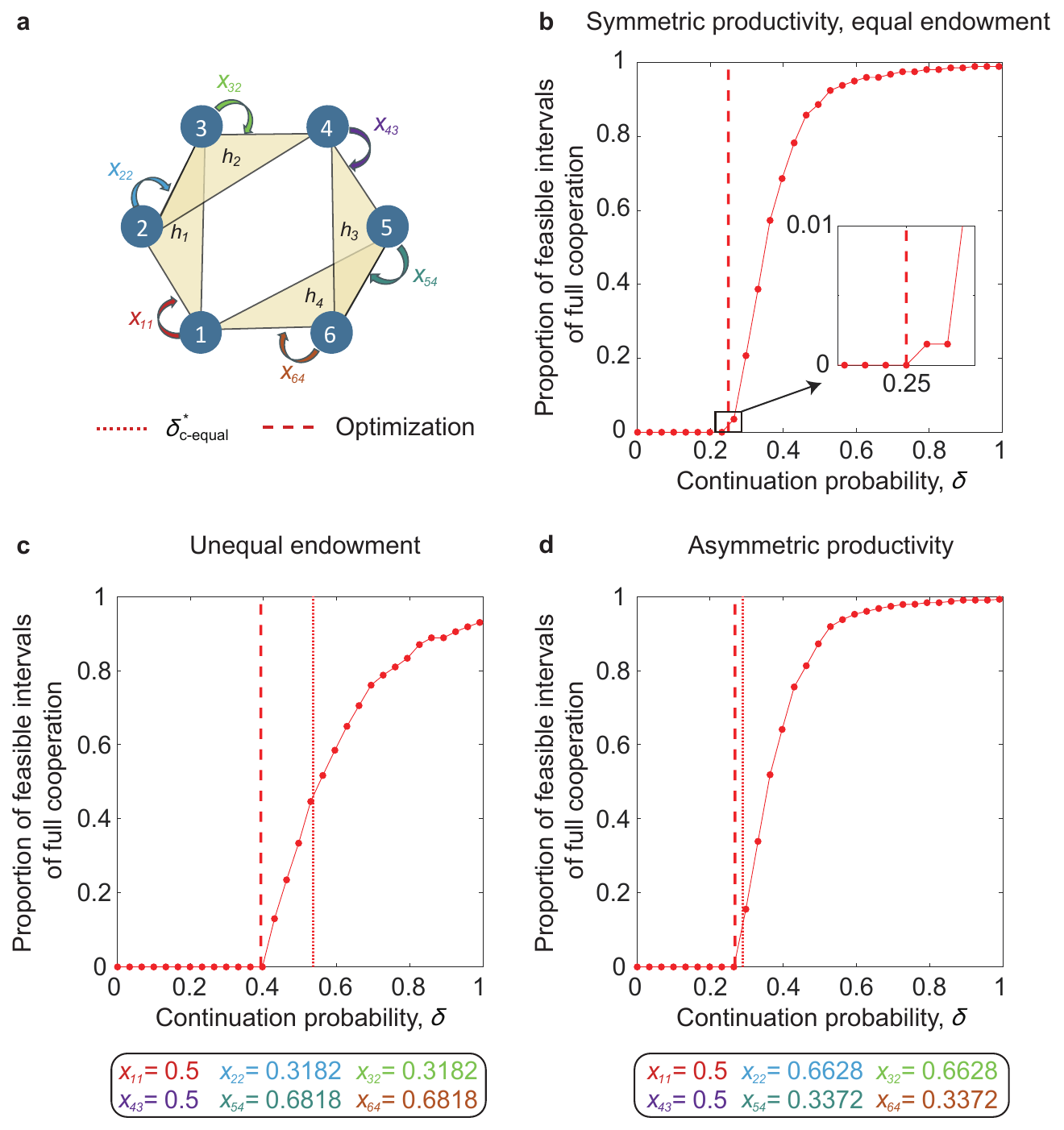}
 \caption{\textbf{Unequal contributions across games enhance cooperation under asymmetric productivity or unequal endowments.} \textbf{a}, A partially connected hypergraph used to identify optimal contribution strategies across the entire contribution space. \textbf{b}, The 
 proportion of fully cooperative combinations under the symmetric productivity and equal endowment, e.g., \(e_i = \frac{1}{6}\) and \(r_i = 2\). The red dashed line represents \(\delta\) for optimized contributions (see the Materials and Methods section), while the red dotted line indicates \(\delta_{\text{c-equal}}^{*}\) for equal contributions. Their alignment indicates that equal contributions best promote cooperation. \textbf{c}, Introducing unequal endowments while maintaining symmetric productivity (\(e_1=0.3\), \(e_{2}=e_{3}=e_{4}=e_{5}=e_{6}=0.14\)) creates a gap between optimized \(\delta\) and equal contributions, suggesting that unequal contributions better facilitate full cooperation. Optimal contributions are highlighted in boxes: Player 1, participating in \(h_1\) and \(h_4\), receives a larger endowment, prompting other players in these hyperedges to contribute more, thereby enhancing cooperation. \textbf{d}, When endowments remain equal but productivity is asymmetric (\(r_1=2.5\), \(r_{2}=r_{3}=r_{4}=r_{5}=r_{6}=1.9\)), a similar gap emerges, demonstrating that unequal contributions are more effective in fostering full cooperation. The optimal strategy in \textbf{d} reverses that of (\textbf{c}): higher productivity for Player 1 means other players in \(h_1\) and \(h_4\) should contribute less to promote full cooperation.} 
    \label{fig4}
\end{figure}

In the above study, we fixed a contribution proportion \( p \) and determined productivity, investigating the proportion of feasible intervals of full cooperation. Below, we discuss the scenario where a fixed endowment distribution and players' productivity are given, examining which contribution strategies among all possible ones are most conducive to promoting cooperation. We continue to use a partially connected homogeneous hypergraph with 6 nodes as our study subject (Fig. \ref{fig4}\textbf{a}).

We find that when productivity is fixed as symmetric and endowments are equal, the optimal contributions are equal contributions (Fig.~\ref{fig4}\textbf{b}). This finding aligns with the conclusions drawn from the partially connected homogeneous hypergraph (Figs.~\ref{fig2} and \ref{fig3}). However, when we introduce unequal endowments by allocating more to player 1 (Fig.~\ref{fig4}\textbf{c}), equal contributions are no longer the optimal strategy. Instead, the optimal contribution strategy requires that the other players connected to player 1 through hyperedges contribute a larger portion of their endowments to those hyperedges.

Next, we explore the scenario where we alter player 1's productivity by increasing it (Fig.~\ref{fig4}\textbf{d}). In this case, we observe that equal contributions are not the optimal strategy. Compared to the previous scenario driven by unequal endowments, the optimal strategy now requires that the other players connected to player 1 through hyperedges contribute less of their endowments to those hyperedges. These two findings correspond directly to our inequality (\ref{linear asymmetric sufficient}), players with large endowments and low productivity are the most likely to deviate from cooperation. Correspondingly, when we increase player 1's endowment, other players need to contribute more to hyperedges involving player 1 to reduce his likelihood of deviation. Conversely, when player 1's productivity is increased, implying that the productivity of other players is relatively lower, there is no need to contribute more to player 1's hyperedges. Instead, contributions should be reduced, thereby increasing the likelihood of full cooperation.

\subsection*{Approaches for promoting full cooperation}

In social interactions, most networks are heterogeneous, underscoring the necessity of studying heterogeneous hypergraphs. Here we introduce two effective approaches for promoting cooperation within small-scale heterogeneous hypergraphs. In the following section, we apply these approaches to large-scale heterogeneous hypergraphs to validate their effectiveness.

\begin{figure}
    \centering
    \includegraphics[width=0.8\linewidth]{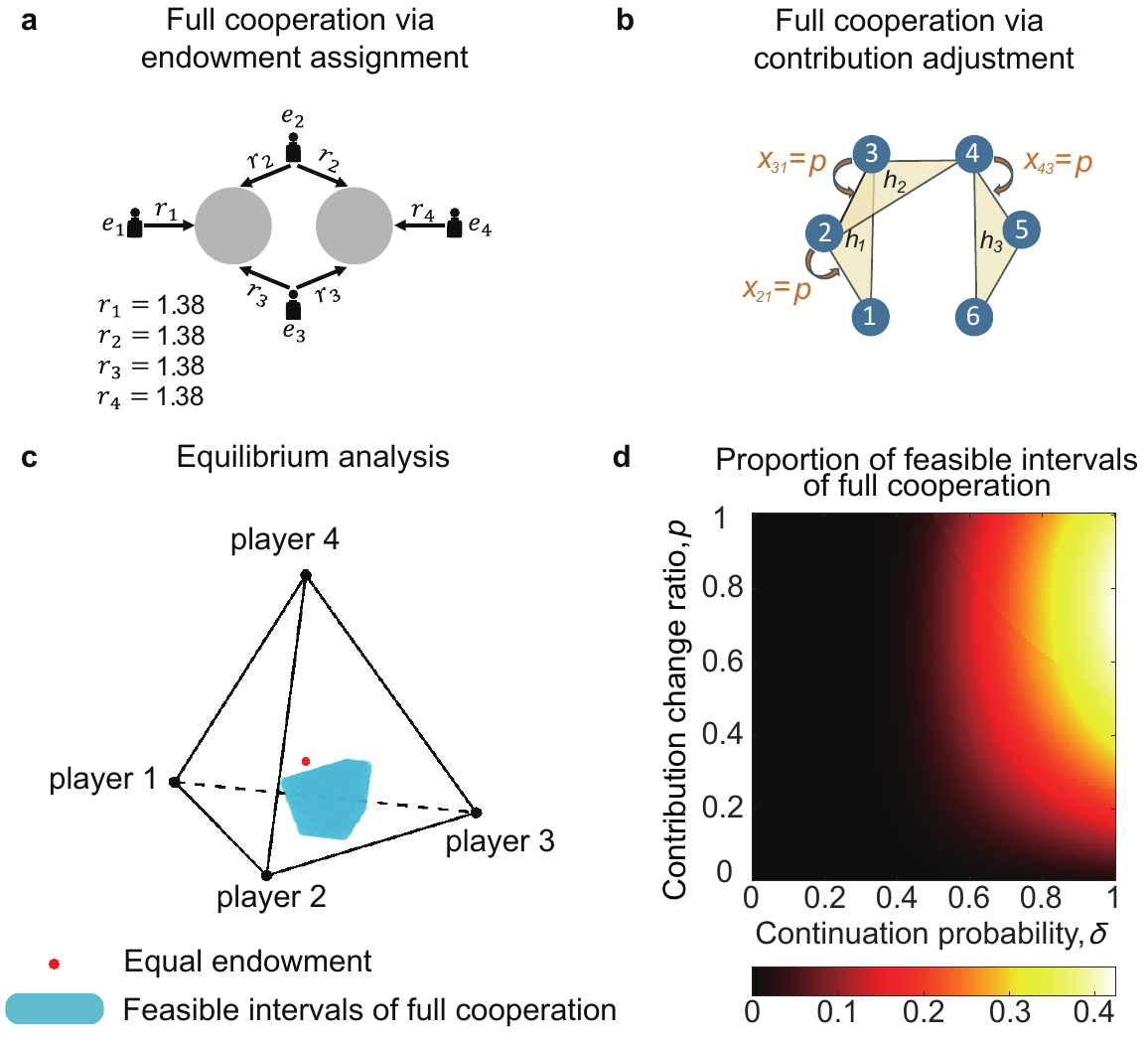}
    \caption{\textbf{Full cooperation on heterogeneous hypergraphs.} We examine two heterogeneous hypergraphs: one with four players (\textbf{a}) and another with six players (\textbf{b}). \textbf{a}, Under a symmetric productivity payoff function, all players have a productivity of \( r = 1.38 \) and a continuation probability of \( \delta = 0.9 \). Players 2 and 3 distribute their endowments equally across both games. \textbf{c}, The endowment space for four players is visualized as a positive tetrahedron, where the equal endowment vector, \( \boldsymbol{e} = \left\{\frac{1}{4}, \frac{1}{4}, \frac{1}{4}, \frac{1}{4}\right\} \), is marked by a red point at the center. The fully cooperative feasible region, shown in blue, reveals that allocating more endowments to players 2 and 3 enhances full cooperation. \textbf{b}, In a six-player heterogeneous hypergraph, players are allowed to adjust their contributions rather than adhering to equal contributions, {with all players’ productivity set to $r_i = 2$}. \textbf{d}, As the contribution proportion $p$ varies from 0 to 1, the proportion of fully cooperative feasible combinations first increases and then decreases, as indicated by the brightness of the color bar. This suggests that equal contributions are no longer optimal. Instead, players with higher degrees should contribute more to hyperedges associated with lower-degree players to enhance cooperation. Overall, cooperation can be effectively promoted through two key mechanisms: allocating more endowments to players involved in a greater number of games and encouraging individuals to allocate their contributions toward hyperedges connected to lower-hyperdegree players.}
    
    \label{fig5}
\end{figure}

In small-scale heterogeneous hypergraphs, we investigate two distinct hypergraph structures. The first is a heterogeneous hypergraph consisting of four players, where players 1, 2, and 3 participate in one game, and players 2, 3, and 4 participate in another game (Fig.~\ref{fig5}\textbf{a}).  Players 2 and 3 allocate their endowments equally between the two games. The second structure involves a heterogeneous hypergraph composed of six players, where players 1, 2, and 3 form one game, players 2, 3, and 4 form a second game, and players 4, 5, and 6 form a third game (Fig.~\ref{fig5}\textbf{b}). Players 2 and 3 contribute a proportion \( p \) to hyperedge \( h_1 \), while player 4 contributes a proportion \( p \) to hyperedge \( h_3 \).

In homogeneous hypergraphs with a symmetric productivity payoff function, equal endowments are most conducive to cooperation. However, this conclusion does not hold in heterogeneous hypergraphs. Under a symmetric productivity payoff function, we observe that when the continuation probability $\delta$ is set to 0.9, the feasible intervals of full cooperation (depicted in blue) do not include the red dot representing equal endowments (Fig.~\ref{fig5}\textbf{c}). Furthermore, we find that full cooperation becomes feasible when more endowments are allocated to players 2 and 3, who participate in more games (Fig.~\ref{fig5}\textbf{c}). Consequently, from the perspective of policymakers, the first approach to promoting full cooperation  is to allocate more endowments to players who participate in a larger number of games.

Changing the distribution of endowments from a collective perspective requires the government or a leader to implement a global allocation strategy. From the perspective of individual players, however, players can only modify how they contribute their own endowments. Therefore, we investigate how players in heterogeneous hypergraphs can adjust their contribution strategies to achieve full cooperation. When we adjust the contribution proportion \( p \) from 0 to 1, we find that at \( p = 0 \), player 1 becomes an isolated node, making full cooperation impossible (Fig.~\ref{fig5}\textbf{d}). As \( p \) increases to 0.5, with players contributing equally, the proportion of full cooperation does not reach its maximum. When \( p \) is further increased to approximately 0.8, the proportion of full cooperation peaks. This occurs because players 1, 5, and 6 each participate in only one game. From the perspective of $\delta_{i}^{*}=\frac{(\sigma-r_i)e_i}{\sum_{k=1}^{M}a_{ik}\sum_{j\neq i}^{N}r_jx_{jk}e_j}$, their denominators are smaller, necessitating that other players involved in their games allocate more endowments to them to facilitate cooperation. In contrast, players 2, 3, and 4 participate in two games, resulting in larger denominators for \( \delta_i^{*} \) and thus providing greater redundancy. However, when \( p \) continues to increase beyond 0.8, the proportion of full cooperation decreases (Fig.~\ref{fig5}\textbf{d}). This is because, although the denominators of \( \delta_i^* \) for players 1, 5, and 6 have increased, the denominators for players 2, 3, and 4 decrease more significantly, causing \( \max\{\delta_2^{*}, \delta_3^{*}, \delta_4^*\} > \max\{\delta_1^*, \delta_5^*, \delta_6^*\} \). Therefore, from the individual's perspective, the second approach to promoting full cooperation is for players to allocate a larger portion of their endowments to hyperedges associated with players who have lower hyperdegrees.

\subsection*{Full cooperation on large hypergraphs}

\begin{figure}
    \centering
    \includegraphics[width=\linewidth]{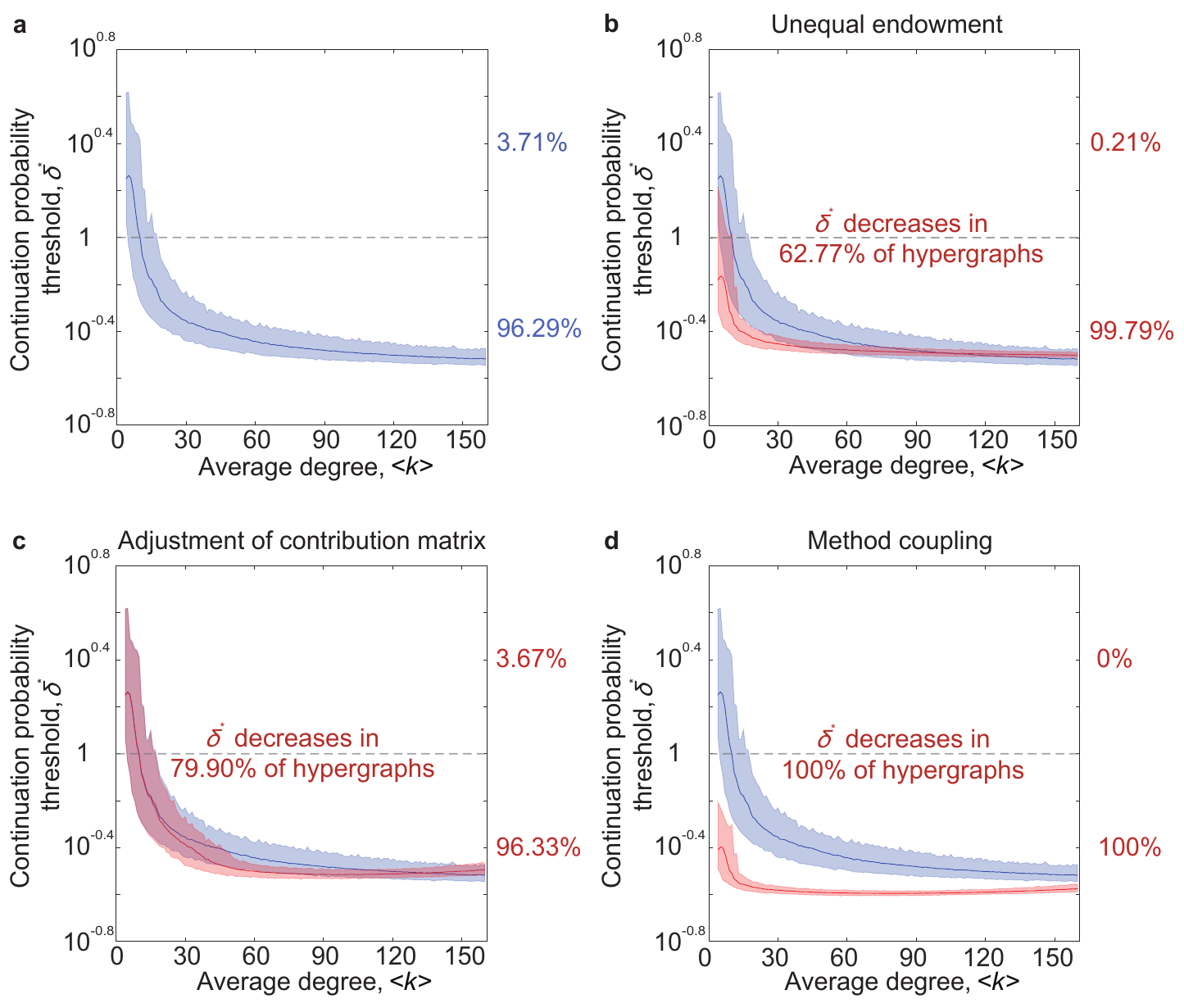}
\caption{\textbf{Enhancing full cooperation in large-scale hypergraphs.} We analyzed 78,500 hypergraphs, with graph size \(N\) randomly selected between 80 and 160 and an average degree $\langle k \rangle$ ranging from 4 to 160.
\textbf{a}, Under symmetric productivity \(r=2\), equal endowments \(\boldsymbol{e} = \left\{\frac{1}{N}, \frac{1}{N}, \dots, \frac{1}{N}\right\}\), and equal contributions, we calculated the critical smallest continuation probability
\(\delta^{*}\). The 95\% confidence intervals for each hyperdegree are shown, with solid lines representing expected values.
On hypergraphs with \(\delta^{*} < 1\), 
full cooperation is feasible for some $\delta>\delta^{*}$, accounting for 96.29\% of cases, whereas for hypergraphs where \(\delta^{*} > 1\), full cooperation is impossible, representing 3.71\%.
\textbf{b}, The effect of adjusting endowments across individuals on the critical continuation probability \(\delta^{*}\). Nodes are divided into two groups: the 25\% with lower hyperdegrees and the 75\% with higher hyperdegrees. We allocate 10\% of the total endowments to the lower-degree group and 90\% to the higher-degree group, distributing endowments evenly within each group. This redistribution increases the proportion of hypergraphs where full cooperation is feasible to 99.79\% and lowers \(\delta^{*}\) in 62.77\% of hypergraphs.
\textbf{c}, The effect of adjusting contributions across different games on \(\delta^{*}\). Contributions are optimized to favor nodes with fewer hyperedges, potentially reducing \(\delta^{*}\). This adjustment increases the proportion of hypergraphs where full cooperation is feasible to 96.33\% and reduces \(\delta^{*}\) in 79.90\% of hypergraphs.
\textbf{d}, The critical continuation probability $\delta^{*}$ by adjusting both endowments to different individuals and contributions to different games.
The combined effect of adjusting both endowments across individuals and contributions across games. This dual optimization ensures full cooperation across all hypergraphs and reduces \(\delta^{*}\) universally.}
    \label{fig6}
\end{figure}

In our investigation of hypergraphs, we initially focus on small structures with no more than six nodes. To extend our analysis to larger systems and more thoroughly evaluate the emergence of full cooperation, we examine two well-established classes of synthetic hypergraphs: Erdős–Rényi (ER) and Barabási–Albert (BA) hypergraphs. The specific procedures for generating these hypergraphs are provided in the Materials and Methods section. The network sizes range from 80 to 160 nodes, with average degrees $\langle k \rangle$ varying from 4 to 160. This range encompasses both sparse hypergraphs, where \(N \gg \langle k \rangle\), and dense hypergraphs, where $\langle k \rangle$ approaches \(N\).

We analyze the critical smallest continuation probability \(\delta^{*}\) across 78,500 Erdős–Rényi hypergraphs under a symmetric productivity payoff function (Fig.~\ref{fig6}). The \(\delta^{*}\) landscape is partitioned into two regions by a gray dashed line: regions where \(\delta^* > 1\), indicating that full cooperation is unachievable, and regions where \(0 < \delta^* < 1\), which define feasible conditions for the emergence of full cooperation. As shown in Fig.~\ref{fig6}\textbf{a}, ER hypergraphs with lower average hyperdegree exhibit higher \(\delta^*\) values, while increasing the average hyperdegree substantially reduces \(\delta^*\). This trend follows from inequality~(\ref{linear asymmetric sufficient}): increasing the average degree raises the denominators of individual terms, thereby reducing each \(\delta_i^{*}\) and, in turn, lowering the overall critical threshold \(\delta^*\).

Building on our earlier proposal of two strategies for promoting full cooperation in small heterogeneous hypergraphs, namely, unequal endowments and modification of the contribution matrix, we extend these approaches to larger hypergraphs to evaluate their effectiveness at scale.
We first examine the transition from equal to unequal endowment distributions. Specifically, nodes are divided into two groups based on their hyperdegrees: the bottom 25\% (low-degree group) and the top 75\% (high-degree group). To introduce inequality, 10\% of the total endowment is allocated to the low-degree group and 90\% to the high-degree group, with resources evenly distributed among nodes within each group. Despite this imbalance, the approach successfully reduces the critical threshold \(\delta^{*}\) in 62.77\% of the hypergraphs (Fig.~\ref{fig6}\textbf{b}).
Next, we consider adjustments to the contribution matrix. Under the baseline equal-contribution scheme, each player \(i\) contributes \(1/k_i\) to each hyperedge they participate in. However, due to degree heterogeneity, players are connected to others with widely varying hyperdegrees. To address this, we preserve the row sums of the contribution matrix \(X\) while slightly modifying its elements: contributions to hyperedges associated with lower-degree nodes are increased by 0.0001, while those to hyperedges linked to higher-degree nodes are decreased by the same amount. This subtle reweighting reduces \(\delta^{*}\) in 79.90\% of the hypergraphs (Fig.~\ref{fig6}\textbf{c}).
When combining these two strategies (unequal endowments and adjusted contributions) we observe a reduction in \(\delta^{*}\) across all hypergraphs, thereby facilitating the emergence of full cooperation (Fig.~\ref{fig6}\textbf{d}). 
{Furthermore, we confirm the generalizability of these findings by extending the analysis to BA hypergraphs (Extended Data~\ref{supplementary_figure2}) and to larger ER hypergraphs with $800–1600$ nodes (Supplementary Fig.~3).}

\subsection*{{Full cooperation in empirical human interaction hypergraphs}}

Building on the above analysis, we have systematically examined the problem of achieving full cooperation in homogeneous, heterogeneous, and large-scale synthetic hypergraphs, and proposed two effective mechanisms to facilitate full cooperation. However, real-world systems exhibit diverse and complex structural characteristics, which may limit the applicability of methods validated in synthetic settings. Therefore, it is crucial to further investigate whether the proposed mechanisms remain effective in real-world scenarios. Additionally, we must also explore whether these mechanisms are inherently present and broadly applicable within real systems.

We validated the effectiveness of the two methods on ten real-world hypergraphs. These hypergraphs include human cooperation hypergraphs and molecular drug interaction hypergraphs. A representative example is the DBLP hypergraph, which captures the collaboration hypergraph among scientists: each node represents a scientist, and each hyperedge corresponds to a co-authored paper. Descriptions of the remaining real-world hypergraphs are provided in the Supplementary Table 1. In Fig.~\ref{fig7}\textbf{a}, we compute the continuation probability threshold $\delta^{*}$ required for full cooperation in real-world hypergraphs under equal endowment and equal contribution settings. Each uppercase label represents a specific real hypergraph, while the suffixes ``-3'', ``-4'', and ``-5'' denote sub-hypergraphs constructed using only the 3-, 4-, and 5-dimension hyperedges, respectively. It is evident that for all real hypergraphs, $\delta^{*} > 1$ is required to achieve full cooperation. This indicates that under baseline conditions, none of the 3-, 4-, or 5-dimension sub-hypergraphs derived from the ten real systems can reach full cooperation. In Fig.~\ref{fig7}\textbf{b}, after applying our two methods, we observe that for all hypergraphs, $\delta^{*} < 1$. This implies that as long as the game is played over a sufficiently long duration, full cooperation becomes possible across all hypergraphs. These results provide further empirical support for the robustness and effectiveness of our proposed methods.

\begin{figure}
    \centering
    \includegraphics[width=0.9\linewidth]{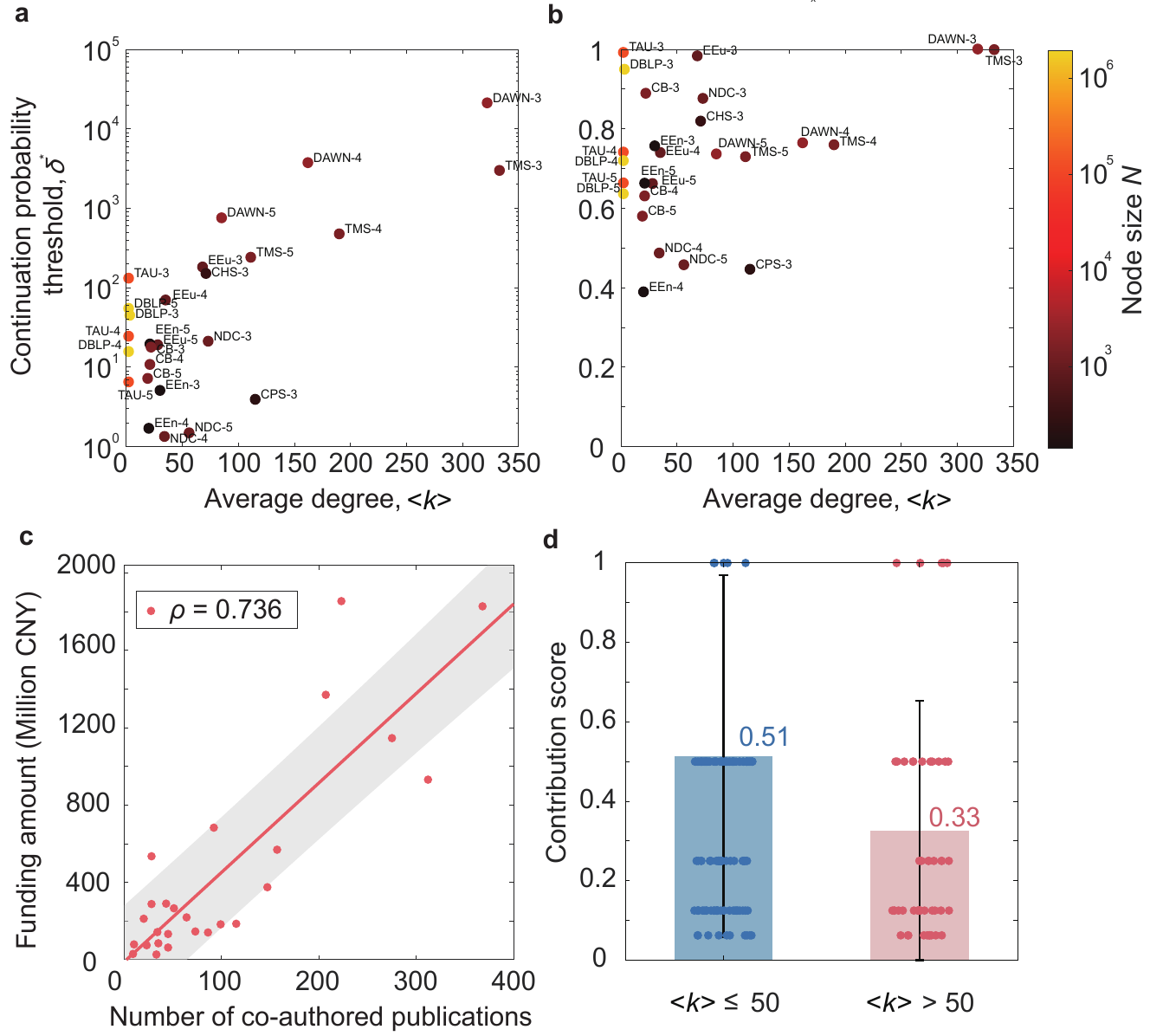}
    \caption{\textbf{Effectiveness of two methods in promoting cooperation on real-world hypergraphs.}
    \textbf{a}, Under equal endowment allocation and equal contribution assignment, we computed the continuation probability threshold $\delta^*$ required for achieving full cooperation. Datasets such as \textbf{DBLP} are labeled using uppercase identifiers, with suffixes \textbf{-3}, \textbf{-4}, and \textbf{-5} denoting sub-hypergraphs constructed from 3-, 4-, and 5-dimension hyperedges, respectively. All empirical hypergraphs yielded $\delta^* > 1$, indicating that full cooperation is not feasible under the baseline setting.
   \textbf{b}, After introducing two mechanisms—unequal endowment and adjustment of contribution—we observed that $\delta^* < 1$ for all empirical hypergraphs, implying that full cooperation becomes achievable.
   \textbf{c}, In the DBLP collaboration hypergraph, we identified 30 scientists and retrieved their corresponding research funding records. The data show that individuals with higher hyperdegree tend to receive more funding, suggesting an empirical bias in resource distribution.
   \textbf{d}, We further examined three highly collaborative scientists and calculated their contributions to other participants within shared hyperedges. Contributions were estimated via an exponentially decreasing scheme based on author order, with corresponding authors contributing a fixed fraction $0.5$ of first-author weight. We found that higher hyperdegree scientists contributed, on average, 51\% to coauthors with a hyperdegree below 50, but only 33\% to those with higher hyperdegrees. The results indicate that higher hyperdegree scientists tend to contribute more to their lowerer hyperdegree coauthors, providing empirical support for both mechanisms within real-world hypergraphs.}
    \label{fig7}
\end{figure}

After validating the effectiveness of the two methods, it is also necessary to examine whether these mechanisms are generally present in real-world systems. To this end, we conduct a case study using the DBLP scientific collaboration hypergraph. The two mechanisms correspond to the following real-world interpretations: (1) social resources (e.g., research funding) are preferentially allocated to higher hyperdegree individuals (i.e., leading scientists), and (2) higher hyperdegree individuals contribute more to lower hyperdegree individuals, akin to mentorship or support for early-career researchers. We begin by verifying the first mechanism—preferential allocation of social resources. In the DBLP hypergraph, we manually collected homepage information for 30 scientists and recorded their total research funding. In Fig.~\ref{fig7}\textbf{c}, the x-axis represents the total number of co-authored papers in 3-, 4-, and 5-person collaborations, and the y-axis represents the total research funding each scientist received. The plot shows that scientists with more publications tend to receive greater research support. Moreover, the relationship between the number of publications and total funding is approximately linear, with a Pearson correlation coefficient of $\rho = 0.736$. This supports the general presence of the first mechanism: social resources are indeed more often allocated to highly connected individuals in scientific collaboration hypergraphs.

The second mechanism concerns whether higher hyperdegree scientists are more likely to support early-career researchers with fewer collaborations. To examine this, we use the order of authorship in scientific papers as a proxy for contribution. In general, the first author typically makes the most significant contribution, while the contribution tends to decrease exponentially with subsequent authors \cite{tscharntke2007author}. Based on this assumption, we assign contribution weights as follows: the first author contributes $100\%$, while the second, third, fourth, and fifth authors contribute $50\%$, $25\%$, $12.5\%$, and $6.25\%$, respectively. Corresponding authors are assigned a fixed contribution of $50\%$. As shown in Supplementary Fig.~1, varying the assumed contribution of the corresponding author does not qualitatively affect the results. We selected three higher hyperdegree scientists from the DBLP hypergraph and collected all their co-authored papers involving 3-, 4-, or 5-person collaborations. For each of these papers, we calculated the average hyperdegree $\langle k \rangle$ of the co-authors (excluding the focal scientist), weighted by the contribution assigned to each author. As shown in Fig.~\ref{fig7}\textbf{d}, the results demonstrate that higher hyperdegree scientists tend to collaborate with and contribute more to lower hyperdegree individuals. This provides empirical support for the second mechanism: highly connected individuals contribute more toward those with fewer connections in real-world collaboration hypergraphs.

\subsection*{{Full cooperation in real-world sustainability domains}}

{The previous subsection primarily demonstrated that two mechanisms capable of promoting full cooperation are widely present in human interaction hypergraphs, with particularly clear evidence observed in scientific collaboration hypergraphs. However, interactions in the real world are not limited to person-to-person relationships. Interactions between countries are equally prevalent and occur across a broad range of sustainability domains. Understanding why and how sustainable cooperation can emerge at the inter-country level is therefore an important question. Here, we show that the two mechanisms proposed in our theoretical framework are consistently observed across multiple sustainability domains. This ubiquity suggests that they may constitute key drivers of global sustainable cooperation and provides theoretical guidance for studying and designing international cooperative systems.}

\begin{figure}
    \centering
    \includegraphics[width=1\linewidth]{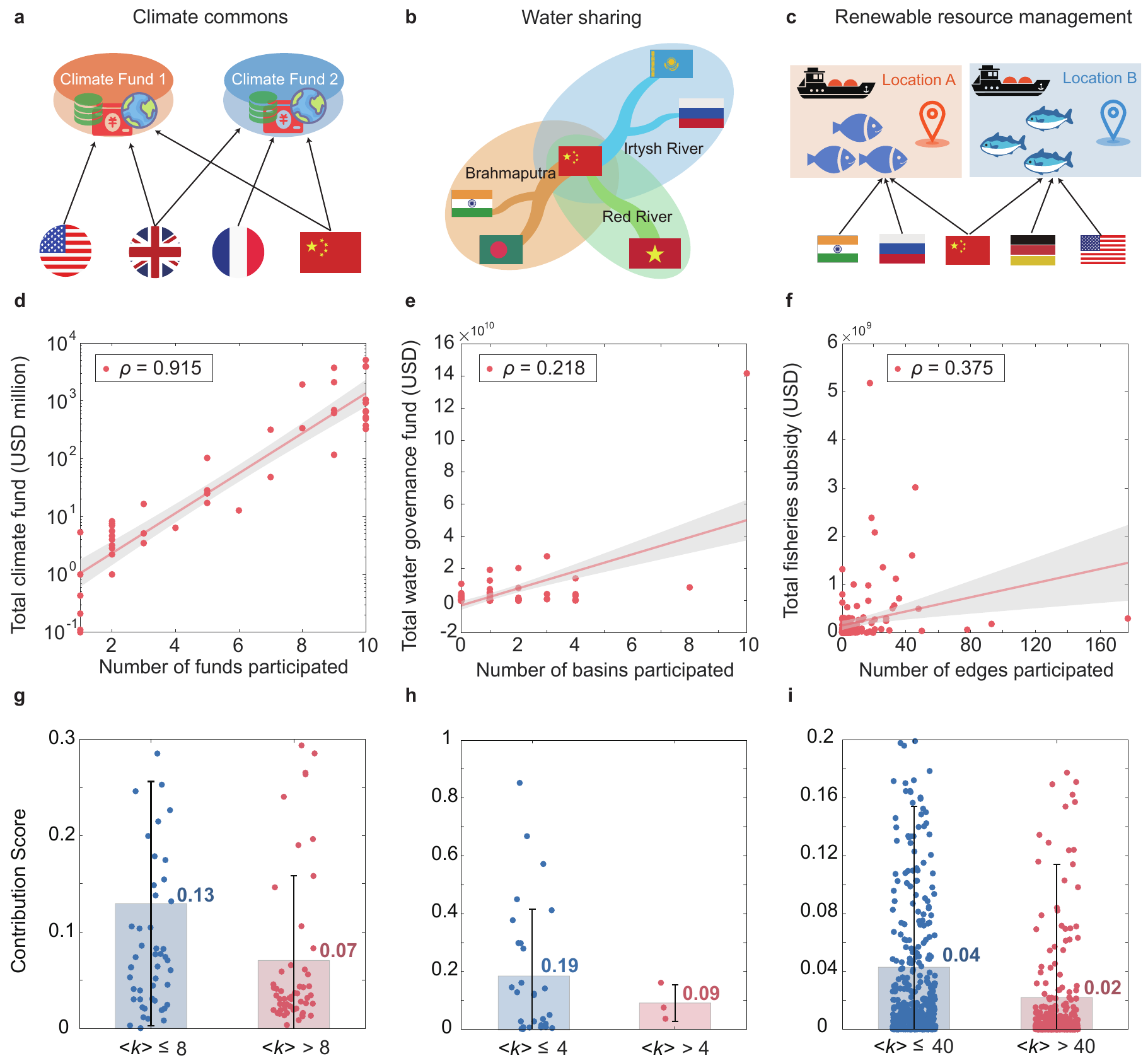}
    \caption{{\textbf{The ubiquity of two mechanisms underlying cooperation across three real-world sustainability domains.} We examine whether the two theoretical mechanisms that facilitate full cooperation generalize to three representative sustainability settings: climate commons (\textbf{a}), water sharing (\textbf{b}), and renewable resource management (\textbf{c}), corresponding to global allocations of national contributions to climate funds, funding allocations for transboundary river governance, and investments for sustaining global fisheries, respectively. Across all three applications, the same theoretical framework applies, with countries as nodes: funds are hyperedges in climate commons, transboundary rivers are hyperedges in water sharing, and fishing grounds are hyperedges in renewable resource management. Using three empirical datasets, we quantify each country’s total endowment (\textbf{d--f}) and its contribution allocation patterns across hyperedges (\textbf{g--i}). We find that total endowment is positively correlated with the number of hyperedges a country participates in, and that high-hyperdegree countries allocate a larger fraction of their endowments to hyperedges with lower average hyperdegree. These empirical regularities align with the two mechanisms predicted by our theory.}}
    \label{fig8}
\end{figure}

\begin{figure}
    \centering
    \includegraphics[width=\linewidth]{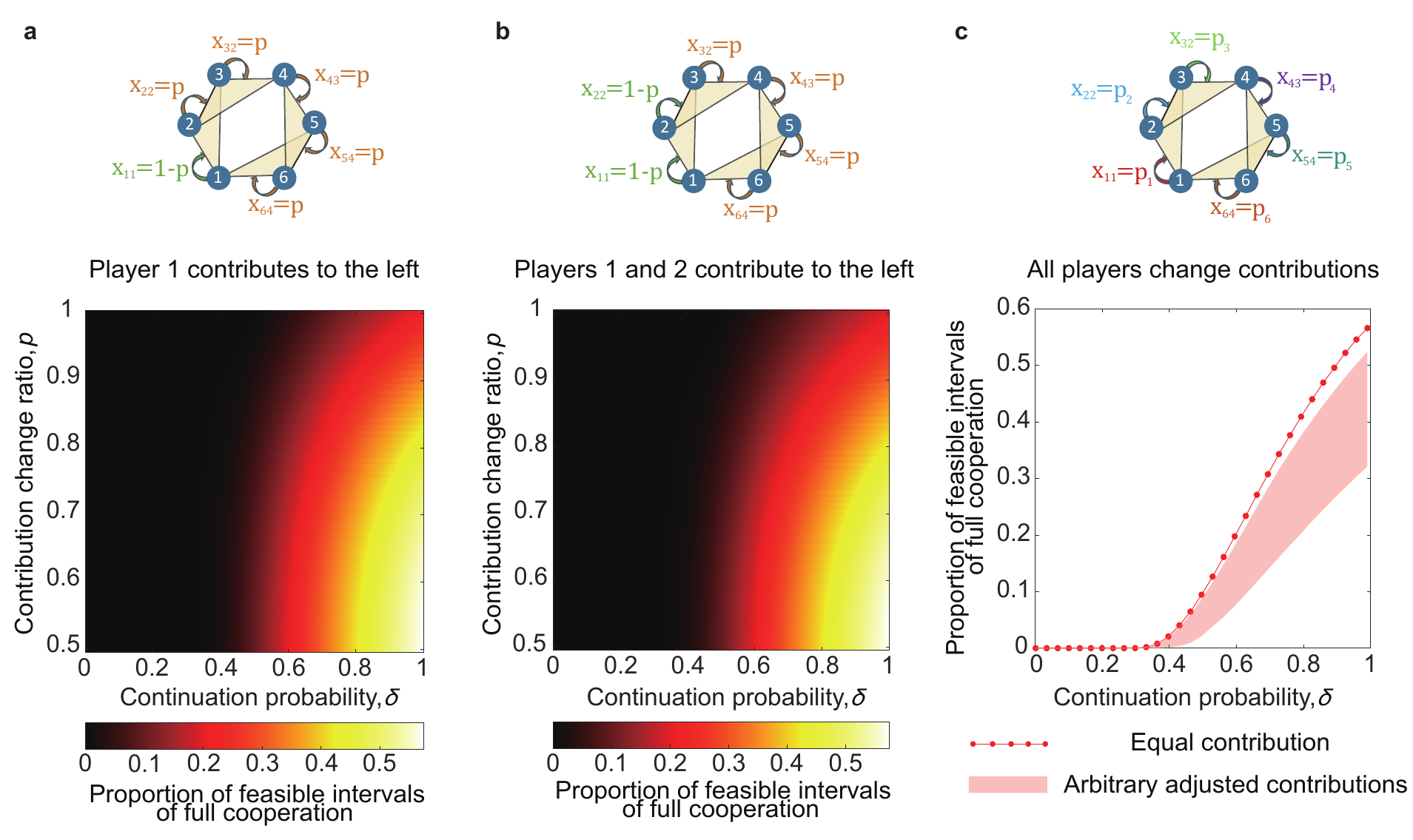}
        \renewcommand{\figurename}{Extended Data}
    \renewcommand{\thefigure}{Fig. 1}
    \caption{\textbf{Equal contributions most effectively promote full cooperation in partially connected homogeneous hypergraphs.} In Figure~\ref{fig3}, we examine how changes in contributions within partially connected homogeneous hypergraphs do not facilitate cooperation, and we identify equal contributions as the most effective for promoting cooperation. While only a limited number of scenarios were discussed in the main text, we expand on this discussion by exploring a wider array of cases here. \textbf{a,} Player 1 has a bias to the left while all remaining players have biases to the right. It is observed that the equal contributions ($p = 0.5$) are most conducive to fostering cooperation. \textbf{b,} Player 1 and player 2 are biased to the left, with the remaining four players biased to the right. Again, the equal contributions ($p = 0.5$) prove to be the most effective in promoting cooperation. \textbf{c,} We allow all players to adjust their biases, and the shaded area represents arbitrary changes in contributions for 100 groups. It is found that the equal contribution consistently exceeds all variations within the shaded regions. Thus, we conclude that the equal contributions are the most favorable for fostering cooperation in homogeneous hypergraphs. {All players’ productivity set to $r_i = 2$}.}
    \label{supplementary_figure1}
\end{figure}

\begin{figure}
    \centering
    \includegraphics[width=\linewidth]{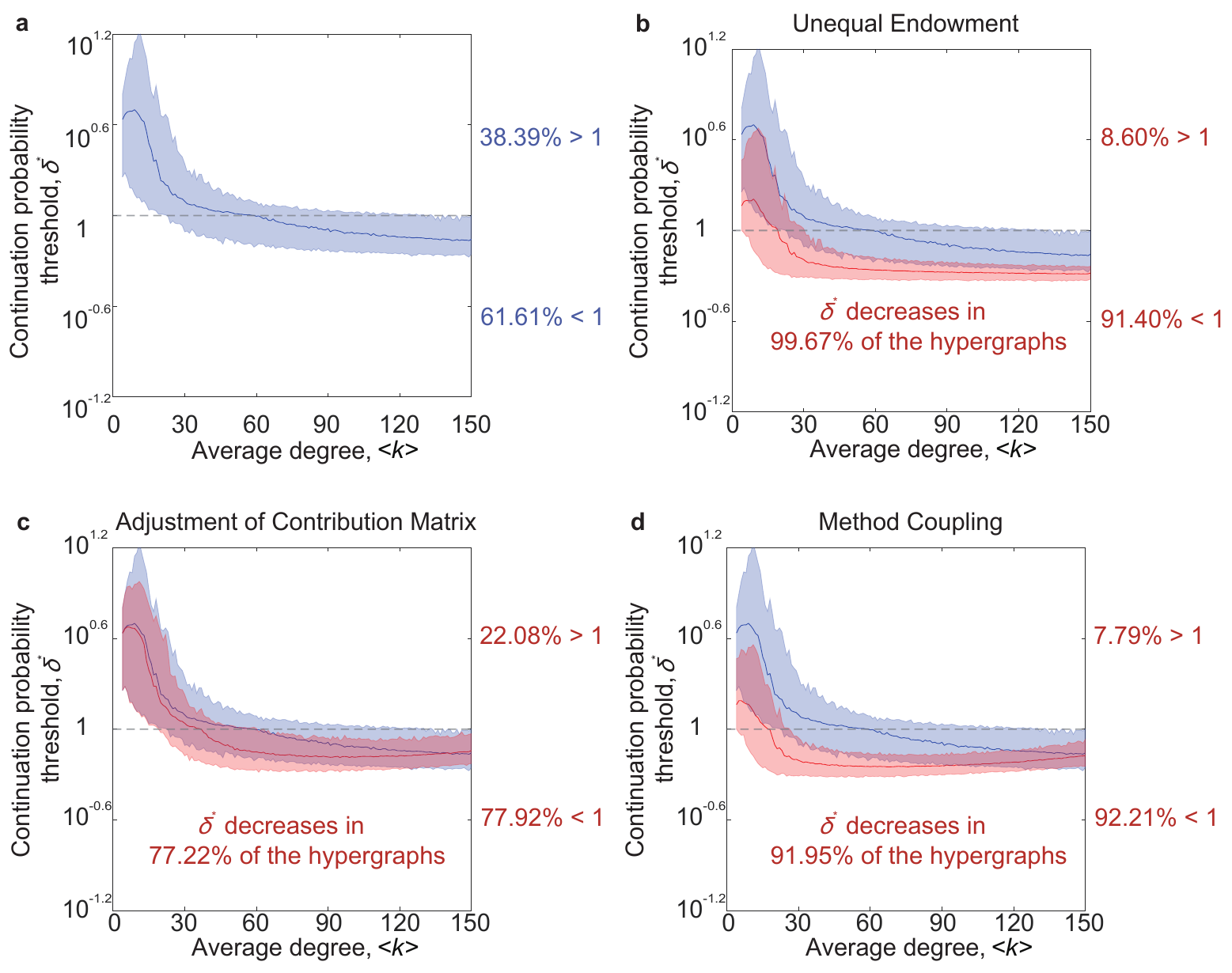}
        \renewcommand{\figurename}{Extended Data}
    \renewcommand{\thefigure}{Fig. 2}
    \caption{\textbf{Enhancing full cooperation in BA hypergraphs.} We analyzed 73,500 BA hypergraphs, each with node counts \(N\) randomly selected between 80 and 160 and average hyperdegree $\langle k \rangle$ ranging from 4 to 150, generating 500 hypergraphs per configuration. 
    \textbf{a,} We calculated the upper bound of \(\delta^{*}\) under symmetric productivity \(r=2\), equal endowment vector \(\boldsymbol{e} = \left\{\frac{1}{N}, \frac{1}{N}, \dots, \frac{1}{N}\right\}\), and equal contributions. We plotted 95\% confidence intervals for each hyperdegree, with solid lines representing expected values.  We categorized the \(\delta^{*}\) intervals into two groups: unable to fully cooperate (\(\delta^{*} > 1\)) and able to fully cooperate (\(\delta^{*} < 1\)).
    \textbf{b,} Unequal endowments: nodes were sorted by hyperdegree into two groups: 25\% with lower and 75\% with higher hyperdegrees. Endowments were divided, with 10\% allocated to the lower group and 90\% to the higher, distributed evenly within each hyperdegree category.
    \textbf{c,} Adjustment of the contribution matrix: we modified contribution ratios to ensure row sums remained at 1. Contributions were tailored to favor nodes with fewer hyperedges, potentially reducing \(\delta^{*}\). The enhanced proportion allocated to less involved nodes was set at 0.0001.
    \textbf{d}, By combining the two methods, \(\delta^{*}\) is reduced in 91.95\% of hypergraphs.
}
    \label{supplementary_figure2}
\end{figure}

{We investigate the existence of these two mechanisms in three representative domains of sustainable cooperation: climate commons, water sharing, and renewable resource management. 
First, in the climate commons context, we examine international climate fund investments (Fig.~\ref{fig8}\textbf{a}), focusing on national contributions to global climate funds recorded in the Climate Funds Update (CFU) database. When applying our model to this setting, countries are treated as nodes and funds as hyperedges. A country’s total investment defines its endowment, while its investment in each fund represents its contribution. The productivity parameter is set to unity.}

{Second, in the context of water resource sharing, we analyse global transboundary river governance (Fig.~\ref{fig8}\textbf{b}) by examining how countries allocate their endowments across shared rivers. In this case, nodes represent countries and hyperedges correspond to transboundary rivers. A country’s total endowment is defined as the sum of a GDP-based allocation and international water-related assistance measured by SDG Indicator~6.a.1. GDP-based allocations are determined by national income levels: low-income countries contribute 0.3\% of GDP, middle-income countries contribute 1\%, and high-income countries contribute 0.5\%. These differentiated contributions reflect structural differences across income groups: low-income countries face severe financial constraints, middle-income countries often require substantial investment in new water infrastructure such as dams, and high-income countries typically incur moderate maintenance costs for already mature water management systems. A country’s contribution to a given transboundary river is proportional to the river’s discharge within that country relative to the total discharge of all transboundary rivers it shares. River discharge data are obtained from the 2018 Transboundary Freshwater Dispute Database (TFDD). Productivity is measured using government effectiveness indicators provided by the World Bank.}

{Finally, in the renewable resource management domain, we consider the sustainable management of global fisheries (Fig.~~\ref{fig8}\textbf{c}). Here, nodes represent countries and hyperedges correspond to fishing grounds. National endowments are derived from the UBC/Sumaila global fisheries subsidy database. A country’s contribution to a fishing ground is defined as its catch from that location divided by its total catch, with catch data obtained from FAO FishStat. Productivity is again measured by national government effectiveness.}

{Across these three sustainability domains, we extract empirical measures of countries’ total endowments and contribution patterns. Regarding total endowments, Fig.~~\ref{fig8}\textbf{d} shows a strong positive correlation between the number of climate funds a country participates in and its total endowment ($\rho = 0.915$), indicating that countries involved in more hyperedges tend to possess larger endowments. Similarly, Fig.~~\ref{fig8}\textbf{e} shows that countries participating in more transboundary rivers have higher total endowments ($\rho = 0.218$), while Fig.~~\ref{fig8}\textbf{f} indicates that participation in a larger number of fishing grounds is associated with higher expenditures required for sustainable fishery recovery ($\rho = 0.375$).}

{We further examine how countries allocate their endowments across hyperedges. In Fig.~~\ref{fig8}\textbf{g}, focusing on countries participating in ten climate fund projects, we find that contributions to hyperedges with an average hyperdegree below eight receive significantly higher contribution scores than those with higher average hyperdegree. The contribution score is defined as the endowment allocated by a country to a given hyperedge divided by its total endowment. Figure~\ref{fig8}\textbf{h} presents a similar analysis for countries involved in more than five transboundary rivers, showing higher contribution scores for rivers with an average hyperdegree below four. Likewise, Fig.~\ref{fig8}\textbf{i} shows that for countries participating in more than fifty fishing grounds, contributions to fishing grounds with an average hyperdegree below forty are significantly higher. We performed robustness checks on the contribution patterns of high-hyperdegree countries and found consistent results (Supplementary Fig.~2). Additionally, we compute the continuation probability thresholds $\delta^{*}$ for three illustrative sustainability domains. We find that $\delta^{*}<1$ for the climate commons, indicating that sustained multilateral engagement of realistic duration can support cooperation. In contrast, the other two domains yield $\delta^{*}>1$, implying that even with the maximum possible persistence of repeated interactions, full cooperation cannot be maintained without modifying the incentive environment.}

{Taken together, these empirical findings provide clear evidence that, across three distinct sustainability domains, the two mechanisms proposed by our theoretical model for promoting full cooperation are robustly supported by real-world data. More broadly, these results offer an empirical foundation for applying our framework to other forms of global cooperation. They suggest two general principles for fostering sustainable cooperation: first, from a system-level perspective, allocating larger endowments to countries with higher hyperdegrees; and second, from an individual-country perspective, directing endowments preferentially towards hyperedges with smaller average hyperdegree.}

\section*{Discussion}

Multi-player interactions are ubiquitous in both natural and social systems, including academic collaborations \cite{ref52}, collective decision-making processes \cite{ref8}, and corporate partnerships \cite{frey2019long}. Moreover, individuals typically do not engage in isolated interactions \cite{ref18,Rand2009Positive}; instead, they participate in multiple group interactions simultaneously \cite{pan2012world,centola2007complex,jaidka2022cross}. To capture this complexity, we introduce a general framework for studying repeated multi-player interactions on hypergraphs. Within this framework, we focus on public goods games, which naturally facilitate the analysis of cooperation under individual heterogeneity. Public goods games not only model how individuals allocate resources within groups but also accommodate various forms of heterogeneity, such as differences in initial endowments \cite{ref16}, productivity levels \cite{ref17}, and contribution strategies \cite{janssen2011coordination,auerbach2013handbook}.
Thus, a comprehensive understanding of how both player-level heterogeneity and the structure of interactions influence cooperation in multi-player settings is crucial for advancing the theoretical foundations of collective cooperation.

\begin{figure}
    \centering
    \includegraphics[width=\linewidth]{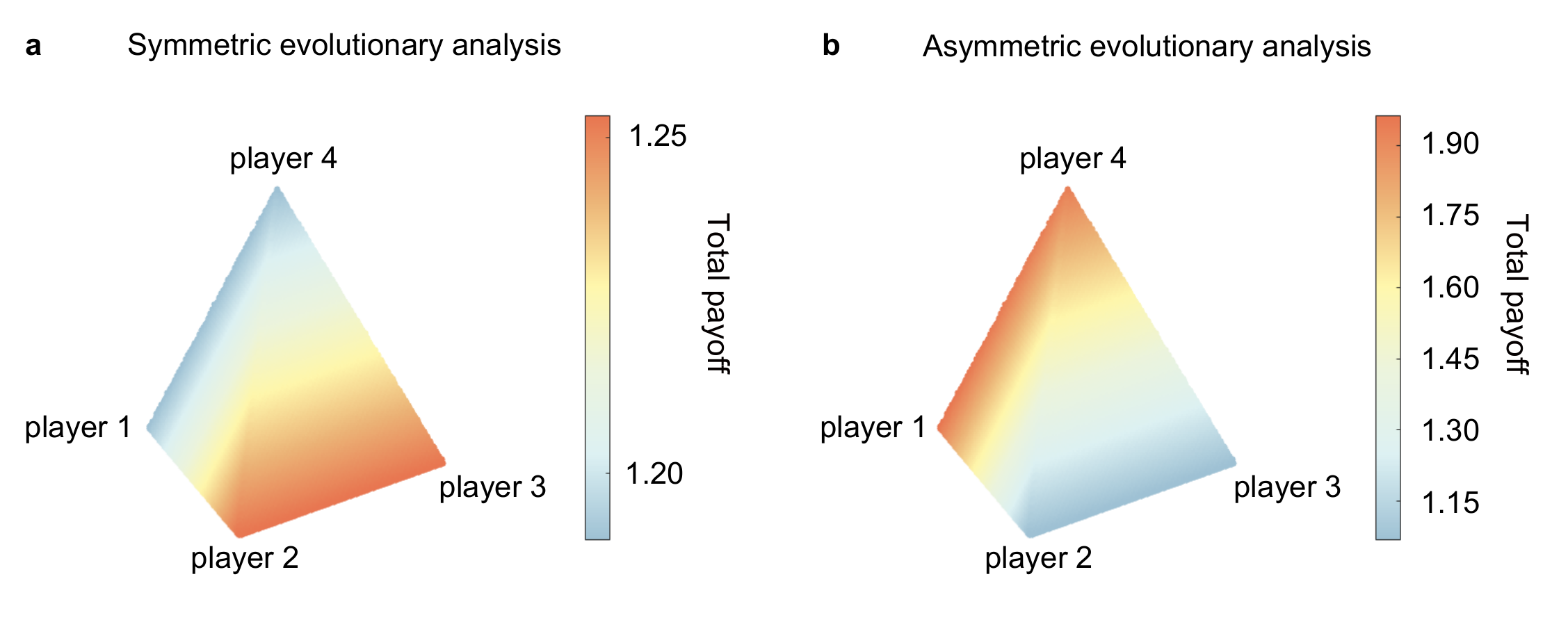}
        \renewcommand{\figurename}{Extended Data}
    \renewcommand{\thefigure}{Fig. 3}
    \caption{\textbf{Evolutionary analysis under linear symmetry and asymmetry.} \textbf{a-b,} We revisit the simplest hypergraph model discussed earlier, where nodes $1$, $2$, and $3$ collectively form a hyperedge $j_1$ and nodes $2$, $3$, and $4$ form another hyperedge $j_2$. The contribution matrix is structured such that at most one element per row is $1$ and all others are $0$. This configuration results in a total state space of $2 \times 3 \times 3 \times 2 = 36$ states. However, not every state is associated with a transition probability, and it is essential to ensure that transitions involve only one strategy change from one state to another. With these definitions and the strategy update rules established, we derive the transition probability matrix $P = \{p_{ij}\} \in \mathbb{R}^{36 \times 36}$. Starting with an initial strategy distribution $\mathbf{v}^0 = \left\{\frac{1}{36}, \dots, \frac{1}{36}\right\}$, we compute the steady state strategy distribution $\mathbf{v} = (1 - \delta) \mathbf{v}^0 (I - \delta P)^{-1}$. Ultimately, the total payoff is calculated as $\Pi = \sum_{i=1}^{36} v_i \times u_i(\boldsymbol{e}, X)$. Parameters are set as follows: (\textbf{a}) linear symmetric payoffs as in Figure~\ref{fig5} ($r_1=r_2=r_3=r_4=1.38,\delta=0.9$), and (\textbf{b}) linear asymmetric payoffs with $r_1=r_4=2.9$, $r_2=r_3=1.1$, and $\delta=0.25$.}
    \label{supplementary_figure3}
\end{figure}

Motivated by the need to understand how heterogeneity and structure influence cooperation, we establish necessary and sufficient conditions for achieving full cooperation in repeated multi-player interactions on arbitrary hypergraph structures. Our approach differs from previous studies, which often focus on fixed probabilities of adopting two or more strategies \cite{ref3} or on the frequency dynamics of strategies \cite{li2018punishment}. Instead, we investigate the problem of endowment allocation under the condition that all participants attain full cooperation in repeated multi-player games on hypergraphs. In a single public goods game, individuals typically allocate their entire endowment to that game, leading to a unique contribution behavior \cite{ref18}. In contrast, achieving full cooperation on hypergraphs requires individuals to distribute their total contributions across multiple simultaneous games, resulting in diverse and non-unique contribution patterns. This flexibility in contribution allocation can significantly affect the likelihood of sustaining full cooperation. Furthermore, we explore the evolution of contribution strategies over a continuous strategy space spanning the closed interval $[0,1]$, as opposed to prior work that primarily considers discrete strategies such as pure cooperation and defection \cite{ref45,ref46}. In our model, players dynamically adjust their contributions through introspective processes \cite{ref18}, continuously evolving their strategies over time. Within the feasible region for full cooperation, this adaptive evolution yields substantially higher payoffs (Extended Data \ref{supplementary_figure3}). To ensure the robustness of our findings, we also perform a sensitivity analysis of the evolutionary parameters (Extended Data \ref{supplementary_figure5}).

\begin{figure}
    \centering
    \includegraphics[width=\linewidth]{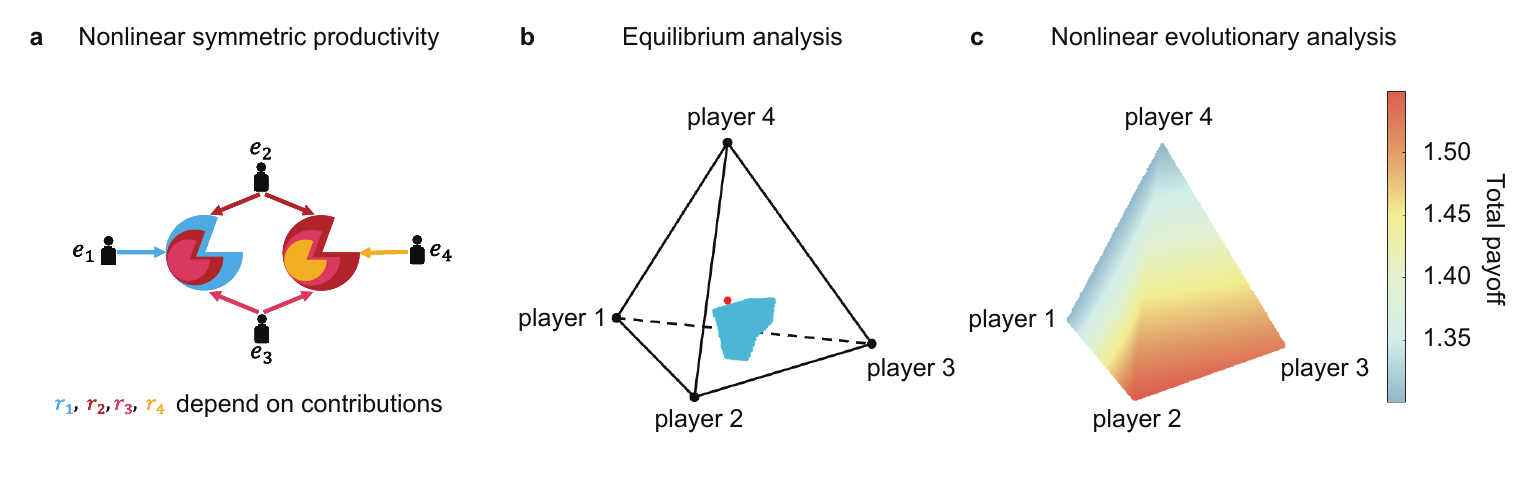}
        \renewcommand{\figurename}{Extended Data}
    \renewcommand{\thefigure}{Fig. 4}
    \caption{\textbf{Equilibrium and evolutionary analysis under nonlinear symmetric payoff function.} \textbf{a,} We investigate nonlinear symmetric payoff functions. We consider a simple hypergraph where players 1, 2, and 3 form one public goods game, and players 2, 3, and 4 form another. \textbf{b,} We delve into feasible intervals of full cooperation for these payoff functions. Here, the set of endowments $\textbf{e}=\{e_1,e_2,e_3,e_4\}$ is conceptualized as a tetrahedron, encapsulating all potential endowment distributions. The light blue region within this tetrahedron demarcates the feasible intervals of full cooperation, with a central red dot symbolizing equal endowments. \textbf{c,} The focus shifts to the variability of total payoff across different endowments under each function. The parameter settings are $\delta=0.65$, $r=1.2$, and $c=0.2$. Initial contributing strategies are preset as $x_{11}=1, x_{21}=0.5, x_{22}=0.5, x_{31}=0.5, x_{32}=0.5, x_{42}=1$.}
    \label{supplementary_figure4}
\end{figure}

\begin{figure}
    \centering
    \includegraphics[width=\linewidth]{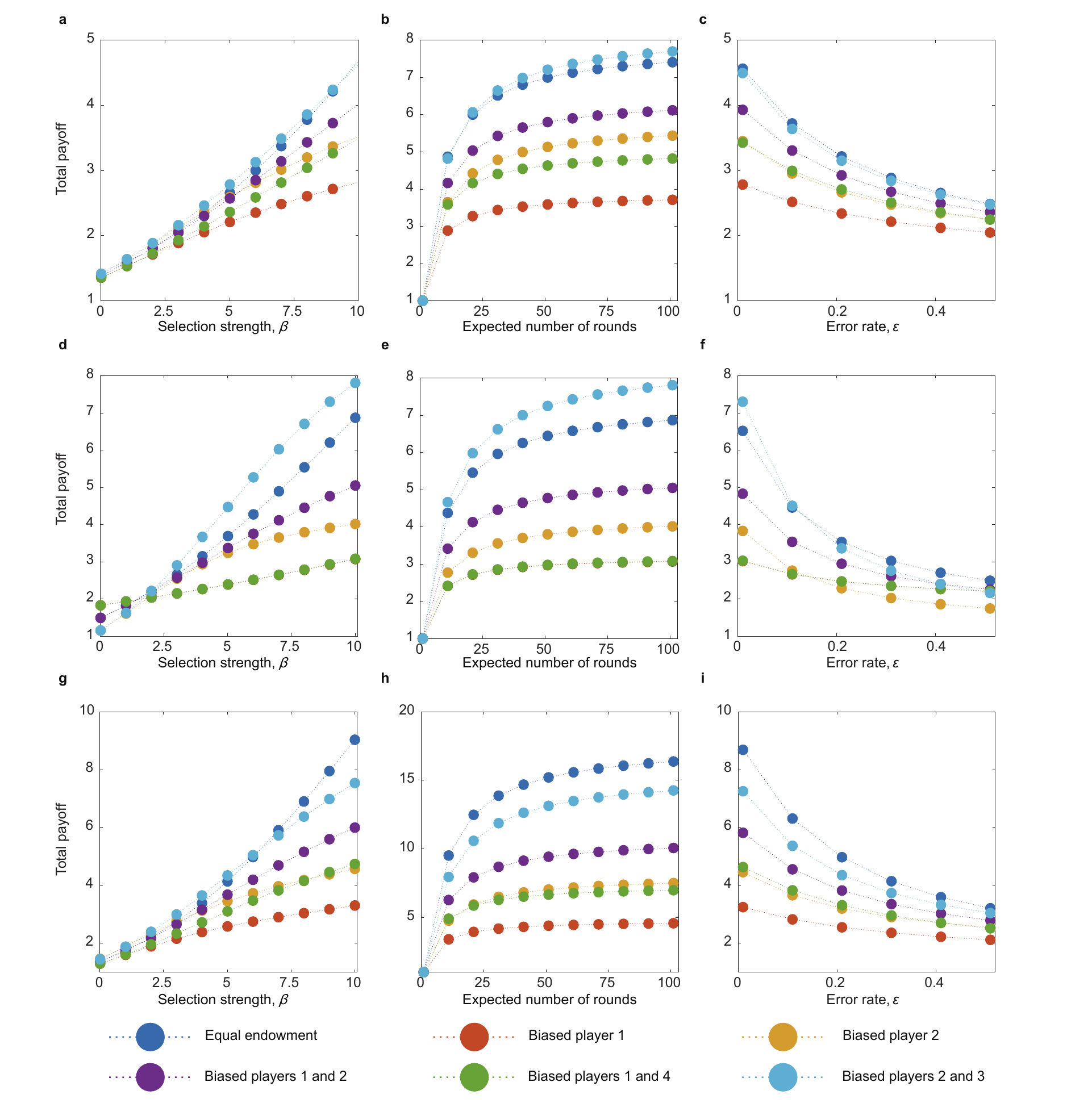}
        \renewcommand{\figurename}{Extended Data}
    \renewcommand{\thefigure}{Fig. 5}
    \caption{\textbf{Robustness of parameters.} We analyze the effects of various factors on total payoffs across different payoff functions and endowment distributions. The hypergraph structure under analysis consists of two hyperedges: one formed by players 1, 2, and 3, and the other by players 2, 3, and 4. \textbf{a-c,} The impacts of coupling strength $\beta$, the evolutionary average round $\frac{1}{1-\delta}$, and the error rate $\epsilon$ are examined using a linear symmetric payoff function as the basis of our analysis. \textbf{d-f,} The linear asymmetric payoff function is explored. \textbf{g-i,} Analyses are conducted under conditions of a nonlinear payoff function. For each analysis, we compare several different endowment distributions, including equal endowments, allocations biased towards player 1, biased towards player 2, biased towards players 1 and 2, biased towards players 1 and 4, and biased towards players 2 and 3.}
    \label{supplementary_figure5}
\end{figure}

\begin{figure}
    \centering
    \includegraphics[width=\linewidth]{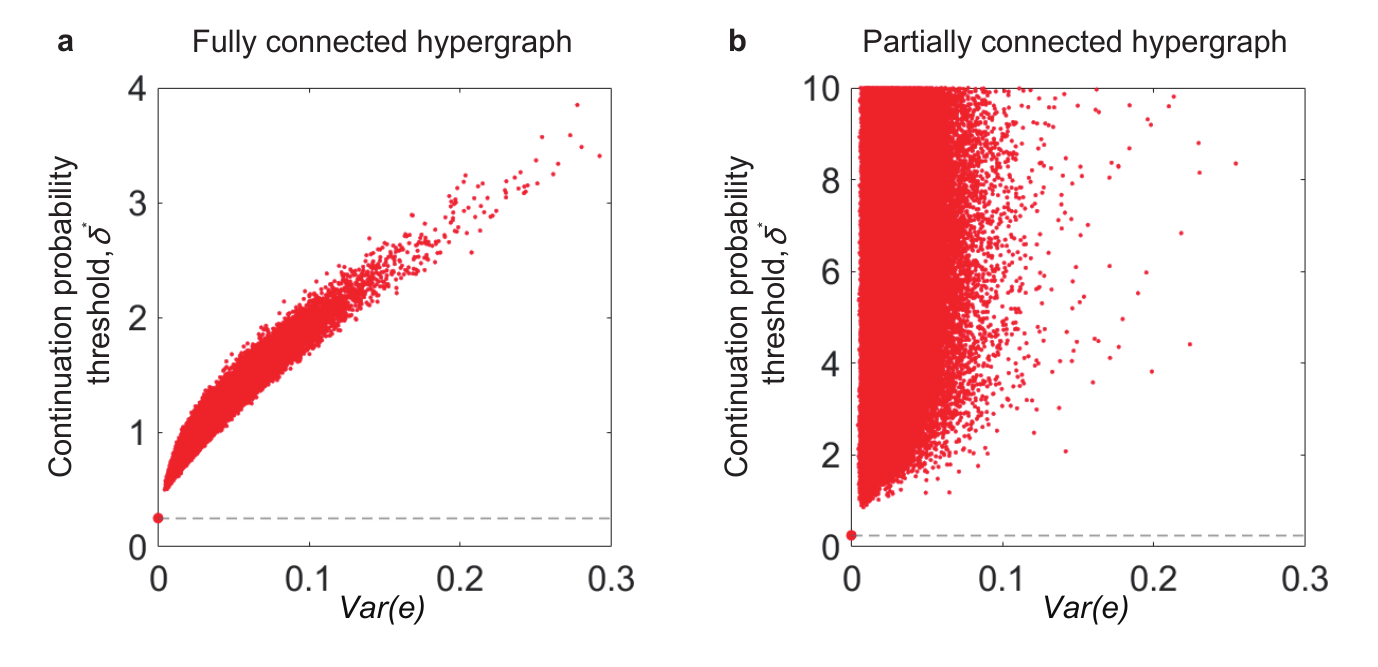}
        \renewcommand{\figurename}{Extended Data}
    \renewcommand{\thefigure}{Fig. 6}
    \caption{\textbf{Verification that equal endowments are most effective for promoting cooperation in large-scale homogeneous hypergraphs.} In Figure~\ref{fig2}, we validated that under symmetric productivity and equal contributions, equal endowments most effectively promote cooperation in small-scale homogeneous hypergraphs with 3-6 nodes. Our theory extends this finding to any homogeneous hypergraph, provided that symmetric productivity and equal contributions are maintained. Therefore, we validated this in large-scale homogeneous hypergraphs with 100 nodes.
\textbf{a}, In a fully connected homogeneous hypergraph with 100 nodes, the x-axis represents the variance of all arbitrary endowments relative to equal endowments. The dashed line indicates the \(\delta^{*}\) required for equal endowments. We found that for all randomly sampled endowments, the corresponding \(\delta^{*}\) values are higher than those for equal endowments.
\textbf{b}, Similarly, in a partially connected homogeneous hypergraph with 100 nodes, we performed the same validation and observed that all randomly sampled endowments correspond to \(\delta^{*}\) values higher than those for equal endowments. Thus, our theoretical results are validated.
}
\label{supplementary_figure6}
\end{figure}

Our work significantly extends the study of Hause et al.~\cite{ref18} by generalizing their single-group model to settings involving multiple interacting groups represented by hypergraphs. Notably, our framework recovers their results as a special case when only a single hyperedge is present, thereby ensuring consistency with the classical single public goods game. In analyzing payoff functions within public goods games, we uncover a critical distinction between games involving a single public good and those structured on hypergraphs. Under conditions of linear symmetry, equal endowment allocations in a single public goods game most effectively promote cooperation. While this principle also holds in homogeneous hypergraphs, we find that in heterogeneous hypergraphs --- even under a linear symmetric payoff structure --- equal endowments do not necessarily maximize cooperation. Instead, our results suggest that full cooperation can be more effectively promoted by allocating larger endowments to individuals who participate in more games.
Further, we extend our analysis to a nonlinear symmetric payoff function, identifying the feasible region for achieving full cooperation (SI Appendix, Section 2.4). Consistent with our earlier findings, we observe that equal endowment allocations are not always optimal for promoting cooperation under nonlinear settings (Extended Data \ref{supplementary_figure4}).

From a policy-making perspective, social resources can be redistributed to promote cooperation. However, at the individual level, players can only adjust their own contribution strategies. Our findings indicate that in homogeneous hypergraphs with symmetric productivity, equal contributions most effectively foster cooperation. Any deviation from equal contributions fails to enhance cooperation. In a homogeneous hypergraph, all players receive an identical share of contributions from others, quantified as $k(\sigma - 1)$. If one player unilaterally adjusts their contribution, some individuals will receive more, while others will receive less. The former group will experience a reduced continuation probability threshold $\delta_i^{*}$, making cooperation easier for them, while the latter will require a higher threshold to maintain cooperation. Since full cooperation requires the continuation probability $\delta$ to satisfy the threshold of every player, any increase in individual threshold demands raises the critical minimum threshold $\delta^{*}$, making cooperation harder to sustain. In contrast, heterogeneous hypergraphs present a different dynamic. Here, players have varying hyperdegrees, leading to differences in contribution capacity. Players with smaller hyperdegrees are more likely to deviate from full cooperation. However, when players with larger hyperdegrees adjust their contributions to support those with fewer connections, it fosters cooperation. This suggests that, to promote cooperation across society, targeted support for individuals with fewer social connections is crucial, as it stabilizes their participation and strengthens overall group cohesion \cite{ref76}.

{Beyond providing two feasible mechanisms to sustain full cooperation in theory, we examine their applicability in three sustainability domains (climate commons, water sharing, and renewable resource management) using real-world data, and find that both mechanisms are consistently aligned with observed contribution patterns across domains. The first mechanism, an endowment-based allocation rule, maps onto widely used equity principles in commons governance and climate justice scholarship \cite{Ostrom1990,Rajamani2006,Okereke2010,PauwMbevaVanAsselt2019}: for example, climate-finance burden-sharing discussions frequently differentiate expectations across countries by economic capacity and responsibility \cite{DellinkEtAl2009,PauwEtAl2022}, and basin-level financing in transboundary water governance often links cost sharing to capacity and negotiated indicators \cite{Schmeier2012,DinarEtAl2007,SadoffGrey2002}. The second mechanism, contribution adjustment, corresponds to adaptive governance and iterative burden rebalancing when heterogeneity creates persistent gaps \cite{Holling1978,Walters1986,FolkeEtAl2005,DietzOstromStern2003}; in practice, this resembles periodically revising contribution shares as information updates about fiscal capacity, exposure, or implementation performance, as well as using matching funds or tiered contribution rules to strengthen compliance with agreed contributions \cite{KarlanList2007,Meier2007,HuckRasul2011}. Importantly, these two mechanisms also clarify an equity--efficiency trade-off \cite{Okun1975}: endowment-based allocation supports perceived fairness and legitimacy, whereas contribution adjustment improves effectiveness by correcting under-provision and stabilizing collective outcomes.}

Multi-player interactions in structured populations have recently attracted considerable attention. However, our work is fundamentally different from these studies in several key aspects.
First, the way we construct multi-player interaction structures differs significantly. Recent research on multi-player games typically starts from pairwise networks, where each node initiates a group interaction involving all its neighbors \cite{ref6, ref47}. Alternatively, some works adopt a set-based representation of games, but still rely on pairwise interactions to compute payoffs \cite{ref62}. These approaches fail to fully capture the complexity inherent in genuine multi-player interactions. 
{Second, the research questions we address are distinct. Most prior studies focus on the evolution of cooperation \cite{ref52} or on identifying critical thresholds that promote cooperation \cite{ref63}. Individuals typically choose between two discrete strategies (cooperate or defect) and apply the same strategy across all interactions. In contrast, we investigate how resource allocation, heterogeneous individual contributions, and environmental conditions jointly shape long-run social welfare. The strategy space is not restricted to discrete choices; instead, it allows any contribution level in the continuous interval $[0,1]$.
Third, our methodology differs markedly. Existing studies predominantly rely on replicator dynamics \cite{ref52}, often assuming bounded rationality or limited foresight. By contrast, our theoretical framework is built on (subgame-perfect) Nash equilibrium. We assume forward-looking, fully rational individuals.
Finally, our conclusions differ both in scope and in underlying mechanisms. While prior work primarily derives structural threshold conditions under which cooperation can emerge \cite{ref52}, our results go beyond predicting the presence of cooperation. We further propose two mechanism-driven design principles that can effectively improve overall social welfare.}

We propose a mathematical framework for analyzing repeated multi-player interactions in structured populations. Despite its broad applicability, several important questions remain open and merit further investigation. These include the emergence of full cooperation under changing game states~\cite{ref78}, as well as the long-term sustainability of cooperation on dynamic higher-order networks~\cite{ref64}. 
While our current focus has been primarily on public goods games, there is substantial potential to extend this framework to other forms of multi-player interactions, such as the Ultimatum Game~\cite{ref65} and multi-player coordination games~\cite{ref66}. Each of these directions presents unique challenges and opportunities for deepening our understanding of strategic behavior in complex social and technological systems. These extensions represent promising avenues for future research.

\section*{Materials and Methods}

We provide a complete mathematical derivation in the Supplementary Information (SI) Appendix. Below, we provide a concise summary of the necessary conditions for achieving full cooperation in our model, describe the optimal solutions illustrated in Figs.~\ref{fig2}, \ref{fig3}, and \ref{fig4}, and detail the two methods employed to construct large-scale hypergraphs.

\subsection*{Equilibrium analysis}

We derive the necessary and sufficient conditions for achieving full cooperation in repeated multi-player games on hypergraph. Specifically, when the payoff function $u$ satisfies Eq.~(\ref{linear_asymmetric_structure}), and given an endowment vector $\boldsymbol{e}$, we establish the following three equivalent conditions (SI Appendix, Lemma 2.1):
\begin{enumerate}
    \item $\bm{e} \in E_u(\delta)$.
    \item For all players $i$ with $e_i > 0$, it holds that\\
        \begin{equation}   \delta(u_i(\boldsymbol{e},\{\mathbf{{1}}_{-i}\})-u_i(\boldsymbol{e},\bm{\textit{O}}))\geq u_i(\boldsymbol{e},\{\mathbf{{1}}_{-i}\})-u_i(\boldsymbol{e},\{\mathbf{1}\}).
        \end{equation}
    \item The strategy profile where all players adopt the Grim strategy constitutes a subgame perfect equilibrium for the given endowment distribution.
\end{enumerate}

By substituting the explicit form of the payoff function $u$ from Eq.~(\ref{linear_asymmetric_structure}) and assuming a given contribution matrix $X$, we derive that the endowment vector $\boldsymbol{e}$ must satisfy the inequality (\ref{linear asymmetric sufficient}) to ensure full cooperation. We then apply this necessary and sufficient condition to specific hypergraph structures. In homogeneous hypergraphs, we rigorously prove that when players have symmetric productivity, equal endowments are optimal for fostering cooperation. However, in heterogeneous hypergraphs, equal endowments are no longer optimal. We provide a counterexample in which a small $\varepsilon$-perturbation around the equal endowments leads to a lower smallest continuation probability, i.e., $\delta^{*} < \delta_{\text{equal}}^{*}$. Therefore, as long as the hypergraph continuation probability $\delta$ satisfies $\delta^{*} < \delta < \delta_{\text{equal}}$, full cooperation can be achieved under the perturbed endowment, while equal endowment fails to do so (SI Appendix, Corollary 1).

\subsection*{Optimal solutions}

In Figs.~\ref{fig2}--\ref{fig4}, we present and compare the corresponding optimal solutions. Figs.~\ref{fig2} and \ref{fig3} share the same optimization objective: identifying endowment distributions that achieve the smallest continuation probability threshold, given fixed productivity and contribution matirx. In contrast, Fig.~\ref{fig4} focuses on optimizing contribution matrix to minimize the continuation threshold under fixed productivity and endowment distributions. Thus, the optimization objectives and constraint settings differ between the two cases.

First, consider the optimization objectives for Figs.~\ref{fig2} and \ref{fig3}:
\begin{equation}
\begin{split}
 \min \quad&  \max_{i} \delta_{i}^{*} = \frac{(\sigma - r_i) e_i}{\sum_{k=1}^{M} a_{ik} \sum_{\substack{j=1, j \neq i}}^{N} r_j x_{jk} e_j}  \\
  \text{s.t.} \quad &\sum_{i=1}^{N}e_i=1.
\end{split}
\end{equation}
Here, \( x_{jk} \) and \( r_j \) are predefined parameters that vary depending on the hypergraph structure. We need to find the optimal endowments.

For Fig.~\ref{fig4}, the optimization objective is defined as:
\begin{equation}
\begin{split}
 \min \quad&  \max_{i} \delta_{i}^{*} = \frac{(\sigma - r_i) e_i}{\sum_{k=1}^{M} a_{ik} \sum_{\substack{j=1, j \neq i}}^{N} r_j x_{jk} e_j}  \\
  \text{s.t.} \quad &x_{11} + x_{14} = 1, \\ 
    &x_{21} + x_{22} = 1, \\ 
    &x_{31} + x_{32} = 1, \\
    &x_{42} + x_{43} = 1, \\
    &x_{53} + x_{54} = 1, \\ 
    &x_{63} + x_{64} = 1.
\end{split}
\end{equation}

In this scenario, \( e_i \) and \( r_j \) are predetermined parameters as specified in Fig.~\ref{fig4}. We need to identify the optimal contributions.

\subsection*{The construction of two large hypergraph structures}

We outline the production process for two hypergraph structures. For the first ER hypergraph, we compute the total number of hyperedges based on random numbers for the production nodes and the average hyperdegree. Subsequently, we assign three nodes randomly to each hyperedge. We then check whether the row sum of the generated adjacency matrix is zero; if it is, the generation process is repeated, and no group is saved.

The second hypergraph is the BA hypergraph, utilized as a model for scale-free static hypergraphs \cite{ref48}. The power-law exponent is set to $\gamma = 2.5$.

\subsection*{Data availability}

{All datasets used in this study are publicly available from the following sources:}

\begin{itemize}
    \item {\textbf{Human-interaction relational networks:} Human-interaction relational datasets were downloaded from \url{https://www.cs.cornell.edu/~arb/data/}.}
    
    \item {\textbf{Climate commons (climate funds):} Climate-fund contribution records were obtained from the Climate Funds Update (CFU) database (\url{https://climatefundsupdate.org/}).}

    \item {\textbf{Water-resource governance (transboundary rivers):} International water-related assistance (SDG Indicator~6.a.1) was obtained from the SDG~6 Data Portal (\url{https://www.sdg6data.org/indicator/6.a.1}) and the UN SDG Global Database (\url{https://unstats.un.org/sdgs/dataportal/}). Transboundary river discharge data were obtained from the Transboundary Freshwater Dispute Database (TFDD) (\url{https://tfddmgmt.github.io/tfdd/index.html}). Government effectiveness data were obtained from the World Bank Worldwide Governance Indicators (WGI) (\url{https://data.worldbank.org/indicator/GE.EST}).}

    \item {\textbf{Renewable resource management (global fisheries):} Fisheries subsidies were obtained from the UBC/Sea Around Us fisheries subsidies database (\url{https://www.seaaroundus.org/fisheries-subsidies/}). Catch data were obtained from FAO FishStat (Global capture production / FishStatJ) (\url{https://www.fao.org/fishery/statistics-query/en/capture}). Government effectiveness data were obtained from the World Bank Worldwide Governance Indicators (WGI) (\url{https://data.worldbank.org/indicator/GE.EST}).}
\end{itemize}

\subsection*{Code availability}

{The code used to generate all results and figures is available at \url{https://github.com/Juyi-Li/Evolution-of-cooperation-on-hypergraphs}.}

\sloppy
\bibliographystyle{unsrtnat}
\bibliography{reference_ns_v1}
\fussy

\section*{Acknowledgements}
\subsection*{Fundings}
Q.S. acknowledges support from the National Natural Science Foundation of China (No. 62473252) and the State Key Laboratory of Autonomous Intelligent Unmanned Systems (No. ZZKF2025-1-4). 

\newpage

\end{document}